\documentclass[twocolumn,final]{svjour3}

\usepackage{amsfonts,amsmath,amssymb}
\usepackage{graphicx}
\usepackage[IT]{subfigure}
\usepackage[usenames,dvipsnames]{color} 			
\usepackage[exponent-product=\cdot,per-mode=symbol]{siunitx} 
\usepackage{ifdraft}

\usepackage[T1]{fontenc} 
\usepackage[utf8]{inputenc} 
\usepackage[swedish,english]{babel} 

\newcommand{\inv}{^{-1}}
\newcommand{\trans}{^{\mathrm{T}}}

\newcommand{\dd}[2]{\frac{\mathrm{d} #1}{\mathrm{d} #2}}

\newcommand{\dede}[2]{\frac{\partial #1}{\partial #2}}

\newcommand{\bp}{\begin{pmatrix}}
\newcommand{\ep}{\end{pmatrix}}
\newcommand{\bbm}{\begin{bmatrix}}
\newcommand{\ebm}{\end{bmatrix}}

\newcommand{\ee}[1]{\cdot 10^{#1}}

\definecolor{LUbronze}{RGB}{156,97,20}
\definecolor{LUblue}{RGB}{0,0,128}
\definecolor{LUlightblue}{RGB}{185,211,220}

\newcommand{\red}[1]{{\color{red}#1}}

\ifdraft{
\newcommand{\todo}[1]{\marginpar{#1}}
\newcommand{\comment}[1]{\red{\emph{\{#1\}}}} 
}{
\newcommand{\todo}[1]{}
\newcommand{\comment}[1]{} 
}

\usepackage{mathptmx}      
\usepackage[authoryear]{natbib} 
\usepackage{url}
\usepackage{empheq}
\usepackage{stackengine} 

\newcommand{\esb}[1]{\begin{subequations} \label{#1} \begin{empheq}[left = \empheqlbrace\,]{align}}
\newcommand{\ese}{\end{empheq}\end{subsequations}}

\newcommand{\pp}[2]{\frac{\partial {#1}}{\partial {#2}}} 

\newcommand{\Vc}{V_{\mathrm{c}}}
\newcommand{\Ag}{A_{\mathrm{g}}}
\newcommand{\rg}{r_{\mathrm{g}}}
\newcommand{\rgtilde}{\tilde{r}_{\mathrm{g}}}
\newcommand{\sg}{s_{\mathrm{g}}}
\newcommand{\sgtilde}{\tilde{s}_{\mathrm{g}}}
\newcommand{\lc}{l_{\mathrm{c}}}
\newcommand{\cb}{\bar{c}}

\newcommand{\lb}{\bar{l}}
\newcommand{\tb}{\bar{t}}
\newcommand{\cc}{c_{\mathrm{c}}}
\newcommand{\Cc}{C_{\mathrm{c}}}

\newcommand{\cs}{c_{\mathrm{s}}}

\newcommand{\ccb}{\bar{c}_{\mathrm{c}}}

\newcommand{\cinf}{\cc^{\infty}}

\newcommand{\lmin}{l^{\mathrm{min}}}

\newcommand{\ct}{\tilde{c}}
\newcommand{\bct}{\vec{\ct}}
\newcommand{\bC}{\vec{C}}

\newcommand{\bA}{\vec{A}}
\newcommand{\bB}{\vec{B}}
\newcommand{\bF}{\vec{F}}

\newcommand{\bI}{\vec{I}}
\newcommand{\bS}{\vec{S}}

\newcommand{\by}{\vec{y}}
\newcommand{\bub}{\vec{\bar{u}}}
\newcommand{\bU}{\vec{U}}

\journalname{Journal of Computational Neuroscience}

\begin{document}\sloppy
\def\stackalignment{l} 

\title{Efficient simulations of tubulin-driven axonal growth
\thanks{E.\ Henningsson was supported by the Swedish Research Council under grant no.\ 621-2011-5588.
}}


\author{Stefan Diehl         \and
        Erik Henningsson		 \and
				Anders Heyden
}


\institute{S.\ Diehl \and E.\ Henningsson \and A.\ Heyden \at
              Centre for Mathematical Sciences, Lund University, P.O.\ Box 118, SE-221 00 Lund, Sweden \\
              \email{diehl@maths.lth.se (S.\ Diehl), erikh@maths.lth.se (E.\ Henningsson, corresponding author), heyden@maths.lth.se (A.\ Heyden).}
}

\date{Received: 29 January 2016 / Revised: 14 March 2016 / Accepted: 5 April 2016 \\
The final publication is available at Springer via http://dx.doi.org/10.1007/s10827-016-0604-x}

\maketitle

\begin{abstract}

This work concerns efficient and reliable numerical simulations of the dynamic behaviour of a moving-boundary model for tubulin-driven axonal growth. The model is nonlinear and consists of a coupled set of a partial differential equation (PDE) and two ordinary differential equations. The PDE is defined on a computational domain with a moving boundary, which is part of the solution. Numerical simulations based on standard explicit time-stepping methods are too time consuming due to the small time steps required for numerical stability. On the other hand standard implicit schemes are too complex due to the nonlinear equations that needs to be solved in each step. Instead, we propose to use the Peaceman--Rachford splitting scheme combined with temporal and spatial scalings of the model. Simulations based on this scheme have shown to be efficient, accurate, and reliable which makes it possible to evaluate the model, e.g.\ its dependency on biological and physical model parameters. These evaluations show among other things that the initial axon growth is very fast, that the active transport is the dominant reason over diffusion for the growth velocity, and that the polymerization rate in the growth cone does not affect the final axon length.

\end{abstract}


\section{Introduction}

We are interested in the modelling of axonal elongation, or growth, from the
stage when one of the developed neurites of the cell body (soma) of a neuron,
begins to grow fast leaving the others behind. The growth can continue for a
long time although with decreasing speed, and axons may also shrink. The main
protein building material of the cytoskeleton consists of tubulin dimers,
which are produced in the soma and transported to the tip of the axon, the
growth cone, in which polymerization of the dimers to microtubules occurs.
This simplified description of the mechanism of the one-dimensional elongation
of the axon has been the focus of both experimental and theoretical works. For
example, the purpose of theoretical work can be to investigate fundamental
questions like the role of advection and diffusion for the transport of tubulin
in long axons without performing tedious experiments. For references on axonal
growth and different types of modelling of the behaviour of the axon and its
growth cone, we refer to the review papers by
\citet{Kiddie2005,Graham2006BMC,Miller2008,vanOoyen2011,Suter2011} and the
references therein.

The dynamic behaviour of a phenomenon is commonly modelled by differential
equations. When an entity, like the concentration of tubulin along the axon,
depends both on time and space, the conservation of mass leads to
{one or several partial differential equations (PDEs)
\citep{Smith2001,McLean2004,Graham2006,Sadegh2010,Garcia2012,SDJTB1}. Tubulin
is in fact present in different states within an axon: motor protein-bound
tubulin and free tubulin. \citet{Smith2001} presented and analyzed an accurate
model of bidirectional transport by motor proteins and free tubulin. These
three states are modelled by three PDEs, two advection equations for the
anterograde (outward from the cell body to the growth cone) and retrograde
(inward) active transports, and one diffusion equation for the movement of free
tubulin. The equations are coupled via reaction terms, or rather
binding/detachment terms, which model the movements of substance between the
free state and either of the actively moving-cargo states. Their model was
successfully calibrated to published experimental data by \citet{Sadegh2010}.

In their publication, \citet{Smith2001} also presented a simplified model of
their three linear PDEs consisting of a single advection-diffusion PDE with
only two lumped model parameters; an effective drift velocity and an effective
drift diffusion constant; see \citet[Formulas (4a)--(4b)]{Smith2001}. It is
such an equation, with an additional sink term modelling the degradation of
tubulin, that was used by \citet{McLean2004,SDJTB1} and which we use in the
present work.}

Since the axon grows, the spatial interval where the tubulin concentration is
defined varies in length and this leads to a moving-boundary problem. Such a
model was presented by \citet{SDJTB1} consisting of a PDE defined on an
interval with moving boundary coupled to two ordinary differential equations
(ODEs). One ODE models the speed of the axon growth, which depends on the
assembly (and disassembly) processes in the growth cone. This ODE was
formulated based on experimental evidence from literature. The assembly process
depends on the available concentration of free tubulin in the growth cone, which
in turn can be modelled by another ODE for the mass balance of tubulin in
the cone. Since this mass balance contains the flux of tubulin along the axon
into the growth cone, the latter ODE is coupled to the PDE. Hence, even for
very simplified assumptions, the mathematical model becomes complicated. All
steady-state solutions were presented in \citet{SDJTB1} and their dependencies on
the values of the biological and physical parameters were investigated. We
refer to that publication for a detailed comparison with previously published
models of axonal growth, in particular, by
\citet{McLean2004,McLean2004num,Graham2006,McLean2006}, since our model can be
seen as an extension of theirs.

It was possible to investigate the dependence on the model parameters of the
steady-state solutions by means of explicit formulas \citep{SDJTB1}.
Furthermore, the stability of each steady state was investigated by numerical
simulations. If a mathematical solution is not stable under disturbances, it is
not physically or biologically relevant and cannot appear in reality. Thus,
while it was possible to describe all steady-state solutions with explicit
formulas, numerical simulation had to be used for dynamic solutions. As was
already noticed by \citet{McLean2004,McLean2004num,Graham2006,SDJTB1}, it is not
straightforward to perform reliable numerical simulations in reasonable CPU
times. The moving boundary can be transformed to a stationary one; however, if
one wants to simulate the outgrowth of an axon from a very small initial length
to its final one, several magnitudes larger, simulations can take months of CPU time
to perform unless a tailored numerical method is used.

It is the main purpose of this article to present an efficient numerical scheme
that can be used for the simulation of the dynamic behaviour of axonal growth.
We also demonstrate the difficulties of using a standard method. Moreover, we
present simulations of the dynamic behaviour of both growth and shrinkage for
variations in the parameters. These simulations give a deeper insight in the
parameters' influence on axonal growth and complement the information from the
steady-state solutions presented in \citet{SDJTB1}.

The efficient numerical scheme presented is obtained by transforming the model
in both space and time, applying a standard second-order spatial
discretization, and using the Peaceman-Rachford splitting time-discretization
\citep{Douglas1955,Peaceman1955,Hundsdorfer,Hansen2013}. Numerical investigations for both the short and long time behaviour indicate the convergence of the numerical
solutions to those of the differential equations, although no proof of
convergence is provided. Furthermore, simulations converge to exact
steady-state solutions when the input soma concentration is constant.

The model equations are reviewed in Sec.~\ref{sec:model} together with the
model parameters. In Sec.~\ref{sec:model_transformation}, the transformations
of the equations in both space and time are given and these are used for the
numerical methods presented in Sec.~\ref{sec:numerical_methods}. Then
Sec.~\ref{sec:simulations} contains several simulations performed partly to
investigate the properties of the numerical methods as such, and partly to
investigate the dynamical properties of the axonal-growth model. Conclusions
are found in Sec.~\ref{sec:conclusions}.

\section{The model} \label{sec:model}

\begin{figure}
\begin{center}
\includegraphics[width=0.4\textwidth]{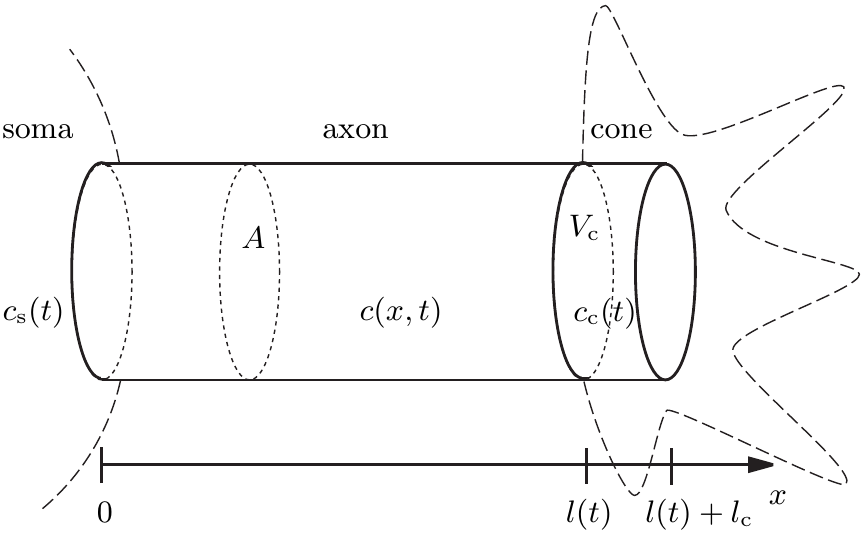}
\end{center}
\caption{Schematic illustration of a growing axon.} \label{fig:Growing_axon}
\end{figure}

An idealized one-dimensional axon is shown in Fig.~\ref{fig:Growing_axon}. The
axon length $l(t)$~[m] at time $t$~[s] is measured from the soma at $x=0$ to
the growth cone. The effective cross-sectional area $A$~[m$^2$] of the axon
through which tubulin is transported is assumed to be constant. Tubulin is
produced in the soma, which is assumed to have the known concentration
$\cs(t)$. This function is the driving input to the model. The unknown
concentration of tubulin along the axon is denoted by $c(x,t)$~[mol$/$m$^3$]
and in the growth cone by $\cc(t)$. {Along the axon, both the motor
protein-bound and the free tubulin are included in $c(x,t)$.} No tubulin is
produced along the axon, but degradation occurs at the constant rate
$g$~[1$/$s]. The active transport by motor proteins is assumed to occur at the
the constant velocity $a$~[m$/$s] and the diffusion of free tubulin is modelled
by Fick's law with a constant diffusion coefficient $D$~[m$^2/$s]. The growth
cone has the volume $\Vc$~[m$^3$]. It turns out that the equations contain the
ratio $\lc:=\Vc/A$, which we therefore interpret as a length parameter
characterizing the size of the growth cone. In the cone, consumption of tubulin
occurs by degradation at the constant rate $g$~[1$/$s] and by assembly of
dimers to microtubules, which elongates the axon at a constant rate
$\rgtilde$~[1$/$s], i.e., $\rgtilde$ is the reaction rate of polymerization of
guanosine triphosphate (GTP) bound tubulin dimers to microtubule bound
guanosine diphosphate (GDP). We let $\Ag$~[m$^2$] denote the constant effective
area of polymerization growth and $\rho$~[mol$/$m$^3$] the density of the
assembled microtubules (the cytoskeleton). Additionally, we assume that the
assembled microtubules in the growth cone may disassemble at the constant rate
$\sgtilde$~[$1/$s]. All biological and physical constants are assumed to be
positive.

The model equations are the following:
\begin{subequations}\label{eq:model}
\begin{empheq}[left = \empheqlbrace\,]{align}
&\dede c t + a \dede c x - D \dede{^2c}{x^2} = -g c, &&\!\!\!\!\!\!\!\!\! 0 < x < l(t),\ t > 0, \label{eq:model1} \\
&\begin{aligned}
\!\dd{\cc}{t}&=\frac{(a-g\lc)}{\lc}\cc-\frac{D}{\lc}c_x^-\\
	&- \frac{(\rg\cc+\rgtilde\lc)(\cc-\cinf)}{\lc}, \\
\end{aligned} &&\!\!\!\!\!\!\!\!\! t > 0,\label{eq:model1b}\\
&\dd l t = \rg(\cc-\cinf), &&\!\!\!\!\!\!\!\!\! t > 0,\label{eq:model1c}\\
&c(0,t) = \cs(t), &&\!\!\!\!\!\!\!\!\! t \geq 0,\\
&c(l(t),t) = \cc(t), &&\!\!\!\!\!\!\!\!\! t > 0,\\
&c(x,0) = c^0(x), &&\!\!\!\!\!\!\!\!\! 0 \leq x < l(0) = l^0,\\
&\cc(0) = c^0(l^0).\label{eq:model1g}
\end{empheq}
\end{subequations}
Equation~\eqref{eq:model1} models the tubulin concentration along the
axon, influenced by advection, diffusion and degradation. {Here we
have used the common assumption that the flux [mol/(m$^2$s)] of tubulin is
\begin{equation}\label{eq:F}
F(c,c_x)=ac-Dc_x,
\end{equation}
where $c_x := \partial c/\partial x$.} The conservation of tubulin in the growth cone is described by
\eqref{eq:model1b}, which { we derive below after motivating
Equation~\eqref{eq:model1c}. The latter equation} states that the growth
velocity due to (net) polymerization is an affine function of the available
concentration $\cc$ in the cone. Since $\cc(t)=\cinf$ is equivalent to
$l'(t)=0$, the constant $\cinf$, appearing both in \eqref{eq:model1b} and
\eqref{eq:model1c}, is the steady-state concentration at which the processes of
assembly and disassembly are equally fast. The background of
Equation~\eqref{eq:model1c} is partly the assumption that the assembly of
tubulin dimers is assumed to be proportional to the amount of tubulin in the
cone $\Vc\cc(t)$ with the reaction rate $\rgtilde$ as the proportionality
constant, and partly that the disassembly occurs at the rate $\sgtilde$ and is
proportional to the amount of already assembled microtubules,
$\rho\Ag\kappa\lc$, where $\kappa>0$ is a dimensionless constant such that
$\kappa\lc$ is the length of the assembled microtubles that may undergo
disassembly. Hence, this disassembly does not depend on the concentration of free tubulin, an assumption in accordance with experiments presented by
\citet{Walker1988}. This leads to the equation
\begin{align}\label{eq:dldt}
&\underbrace{\dd{(\rho\Ag l)}{t}}_{\text{mass increase per unit time}}
=\underbrace{\rgtilde\Vc\cc}_{\text{assembly}}
-\underbrace{\sgtilde\rho\Ag\kappa\lc}_{\text{disassembly}}.
\end{align}
which can be written as
\begin{align}\label{eq:dldt2}
&\dd{l}{t}=\rg\cc-\sg\quad\text{with}\quad
\rg:=\frac{\rgtilde\Vc}{\rho\Ag}\quad\text{and}\quad\sg:=\sgtilde\kappa\lc,
\end{align}
Here $\sg$ is the maximum speed of shrinkage, which occurs when $\cc=0$. The
lumped parameter $\rg$ is a concentration-rate constant. To convert
\eqref{eq:dldt2} to \eqref{eq:model1c}, we define the constant
\begin{align}\label{eq:cinf}
&\cinf:=\frac{\sg}{\rg}=\frac{\sgtilde\rho\Ag\kappa\lc}{\rgtilde\Vc}=\frac{\sgtilde\rho\Ag\kappa}{\rgtilde
A}.
\end{align}

{Equation~\eqref{eq:model1b} originates from the conservation of
mass of tubulin in the growth cone $\cc$:
\begin{equation}\label{eq:cons_law_cone}
\begin{split}
&\underbrace{\dd{(\Vc\cc)}{t}}_{\text{mass increase per unit time}}=\\
&=\underbrace{A\big(ac^--Dc_x^--l'c^-\big)}_{\text{flux in}}
-\underbrace{g\Vc\cc}_{\text{degradation}}
-\underbrace{\rgtilde\Vc\cc}_{\text{assembly}}
+\underbrace{\sgtilde\rho\Ag\kappa\lc}_{\text{disassembly}}.
\end{split}
\end{equation}
The assembly and disassembly terms here are the same as in \eqref{eq:dldt},
however, with opposite signs. The flux [mol$/$s] of tubulin into the growth
cone is the product of $A$, the concentration just to the left of $x=l(t)$,
which is
\begin{align}\label{eq:c-}
c^-& = c^-(t):=c\big(l(t)^-,t\big)
=\lim_{\epsilon\searrow 0}c\big(l(t)-\epsilon,t\big),
\end{align}
and the net velocity of tubulin across $x=l(t)$. The velocity due to advection
and diffusion is $F(c^-,c_x^-)/c^-$ relative the axon, where $F$ is defined in
\eqref{eq:F} and $c_x^-$ is defined in the similar way as \eqref{eq:c-}. Since
the axon is elongated with the speed $l'(t)$, the net flux across the moving
boundary $x=l(t)$ is
\begin{align*}
c^-\left(\frac{F(c^-,c_x^-)}{c^-}-l'\right)
&=F(c^-,c_x^-)-l'c^-\\
&=ac^--Dc_x^--l'c^-,
\end{align*}
which explains the flux term of \eqref{eq:cons_law_cone}. That equation can now be
rewritten by dividing by $\Vc=A\lc$, and using \eqref{eq:model1c} and
\eqref{eq:cinf} to obtain \eqref{eq:model1b}.}

Initial data are denoted by the super index 0. In the derivations of the model
equations \eqref{eq:model1}--\eqref{eq:model1g}, we have used the natural
assumption that the concentration of tubulin is continuous in space and time.

The parameter values used are shown in Table~\ref{tab:1}. The nominal values
are extracted carefully from the biological literature and we refer to
\citet{SDJTB1} for references and explanations on how the parameter values were
found. In particular, the nominal values for $\rg$ and $\sg$ were calculated
from the experiments reported by \citet{Walker1988}. The exception is the
polymerization reaction rate constant $\rgtilde$ in \eqref{eq:dldt}, for which
we made a qualified guess. This variable does not influence any steady-state
solution and for the dynamic behaviour presented in Sec.~\ref{sec:simulations},
we investigate a wide range of values and can conclude that the nominal value
seems to be reasonable.

\begin{table}[tbh]
\caption{Parameter values.}\medskip

\begin{tabular}{llll}
Parameter & Nominal value & Interval & Unit \\
\hline
$a$ & $1$ & $0.5$--$3.0$ & $10^{-8}$~m$/$s \\
$D$ & $10$ & $1$--$25$ & $10^{-12}$~m$^2/$s \\
$g$ & $5$ & $2.5$--$40$ & $10^{-7}$~s$^{-1}$ \\
$\lc$ & $4$ & $1$--$1000$ & $10^{-6}$~m \\
$\sg$ & $2.121$ & $0.5$--$4.0$ & $10^{-7}$~m$/$s \\
$\rg$ & $1.783$ & $0.9$--$7.2$ & $10^{-5}$~m$^4/($mol\,s$)$ \\
$\rgtilde$ & 0.053 & $0.015$--$0.240$ & s$^{-1}$ \\
$\cinf$ & $11.90$ & $2.80$--$23.57$ & $10^{-3}$~mol$/$m$^3$ \\
$\cs(t)$ & --- & $5.95$--$23.80$ & $10^{-3}$~mol$/$m$^3$ \\
 \end{tabular}
\label{tab:1}
\end{table}

\section{Model transformation}
\label{sec:model_transformation}

As a first step in the construction of an efficient numerical method we scale
the model \eqref{eq:model} both in space and time. The former scaling allows us
to work in a constant spatial domain in contrast to the varying domain defined
by \eqref{eq:model}. The latter scaling grants a time adaptivity which is well
needed due to the huge differences in axon growth rates occurring during
simulations, cf.\ Sec.~\ref{sec:dynamical_properties}.

\subsection{Scaling in space}
\label{sec:scaling_in_space}

As the axon grows (or shrinks) the domain of the PDE \eqref{eq:model1} expands
(or contracts). Thus, straightforward application of an off-the-shelf numerical
method is not possible. This issue is considered by \citet{McLean2004} who made
a spatial scaling transforming the domain of the PDE into the constant interval
$(0,1)$. As a consequence the same number of spatial computational cells can be
used along the axon regardless of its length. Compare also with \citet{SDJTB1,Graham2006}, where numerical computations are performed using this
technique. The spatial scaling is the following:
\begin{align*}
&y:=\frac{x}{l(t)},\quad \pp{y}{x}=\frac{1}{l(t)},\quad
\pp{y}{t}=-\frac{xl'(t)}{l(t)^2}=-\frac{yl'(t)}{l(t)},
\end{align*}
where $x\in [0,l(t)]$ and thus $y\in [0,1]$. With $\cb(y,t):=c(yl(t),t)$, the derivatives can be written as
\begin{align*}
&\pp{c}{x}=\frac{1}{l(t)}\pp{\cb}{y},\quad
\frac{\partial^2c}{\partial x^2}=\frac{1}{l(t)^2}\frac{\partial^2\cb}{\partial y^2},\quad
\pp{c}{t}=\pp{\cb}{t}-\frac{yl'(t)}{l(t)}\pp{\cb}{y}.
\end{align*}
Substituting these into the equations and noting that
\begin{align*}
\frac{a-yl'(t)}{l(t)}=\frac{a-y\rg\big(\cc(t)-\cinf\big)}{l(t)},
\end{align*}
we can write the transformed dynamic model \eqref{eq:model} as
\begin{subequations}\label{eq:modelyonly}
\begin{empheq}[left = \empheqlbrace\,]{align}
&\pp{\cb}{t}+\left(a-y\rg(\cc-\cinf)\right)\frac{1}{l}\pp{\cb}{y} - D\frac{1}{l^2}\frac{\partial^2\cb}{\partial y^2} = -g\cb, \label{eq:modelyonly1}\\
&\begin{aligned}
\!\dd{\cc}{t}&=\frac{(a-g\lc)}{\lc}\cc-\frac{D}{\lc}\frac{1}{l}\cb_y^-\\
	&- \frac{(\rg\cc+\rgtilde\lc)(\cc-\cinf)}{\lc},
\end{aligned}\\
&\dd{l}{t}=\rg(\cc-\cinf),\\
&\cb(0,t) = \cs(t),\\
&\cb(1,t)= \cc(t),\\
&\cb(y,0) = c^0(yl^0),\\
&\cc(0)= c^0(l^0),\\
&l(0) = l^0,
\end{empheq}
\end{subequations}
for $y \in (0,1)$ and $t > 0$.

\subsection{Scaling in time and space}
\label{sec:scaling_in_time}

For short axon lengths the advection and
diffusion effects are large in relation to the domain. This is reflected by the
coefficients $1/l$ and $1/l^2$ in \eqref{eq:modelyonly1}. Thus the model can be
expected to be substantially more difficult to simulate when the axon length is
small. Additionally, recall that the axon length is expected to change multiple
orders of magnitude during its growth. Therefore we would expect that
considerably longer time steps can be taken when $l$ is large compared to the
short steps needed to resolve the fast evolution when $l$ is small. Our
temporal scaling is designed to implement such a desired time adaptivity.

First, we make the assumption that $l(t)>0$ for all $t\ge 0$, define the
dimensionless function
\begin{align} \label{eq:gamma}
&\Gamma(t):=a\int\limits_{0}^{t}\frac{\mathrm{d}s}{l(s)}
\end{align}
and introduce the following coordinate transformation, to be applied on
\eqref{eq:model}:
\begin{align}\label{eq:ysofxt}
\left\{
\begin{aligned}
&y:=\frac{x}{l(t)},\\
&\tau:=\Gamma(t),
\end{aligned}\qquad 0\le x\le l(t),\, t\ge 0.
\right.
\end{align}
Since $\Gamma'(t)=a/l(t)>0$ for all $t\ge 0$, the inverse of $\Gamma$ exists, and \eqref{eq:ysofxt} is equivalent to
\begin{align}\label{eq:xtofys}
\left\{
\begin{aligned}
&x=y\lb(\tau),\\
&t=\Gamma^{-1}(\tau),
\end{aligned}\qquad 0\le y\le 1,\, \tau\ge 0,
\right.
\end{align}
where $\lb(\tau):=l(\Gamma^{-1}(\tau))=l(t)$. It is convenient to introduce the
notation $\bar{t}$ for $\Gamma^{-1}$, i.e.\ $t = \bar{t}(\tau) :=
\Gamma\inv(\tau)$. Furthermore, differentiating the identity $\Gamma(\bar
t(\tau)) = \tau$ gives $\Gamma'(\bar t(\tau))\bar t'(\tau) = 1$ which results
in an ODE to update the original time:
\[ \frac{\mathrm{d}\bar{t}}{\mathrm{d}\tau} = \frac{1}{\Gamma'(\bar{t}(\tau))} = \frac{1}{a/l(\bar{t}(\tau))} = \frac{\lb(\tau)}{a}. \]
We append this ODE to the dynamical system~\eqref{eq:model}. Furthermore, we
set $\ccb(\tau):=\cc(t)$, and redefine $\cb(y,\tau):=c(x,t)$. We have
\[ \left\{ \begin{alignedat}2
&\pp{\tau}{t}=\Gamma'(t)=\frac{a}{l(t)}=\frac{a}{\lb(\tau)},&\qquad&
\pp{\tau}{x}=0,\\
&\cc'(t)=\ccb'(\tau)\dede{\tau}{t}=\frac{a\ccb'(\tau)}{\lb(\tau)},&\qquad&
l'(t)
=\frac{a\lb'(\tau)}{\lb(\tau)},\\
&\pp{y}{t}=-\frac{xl'(t)}{l(t)^2}
=-\frac{ay\lb'(\tau)}{\lb(\tau)^2},&\qquad&
\pp{y}{x}=\frac{1}{l(t)}=\frac{1}{\lb(\tau)}.
\end{alignedat} \right. \]
Furthermore,
\[ \left\{ \begin{aligned}
&\pp{c}{x}=\frac{1}{\lb(\tau)}\pp{\cb}{y},\qquad
\frac{\partial^2c}{\partial x^2}=\frac{1}{\lb(\tau)^2}\frac{\partial^2\cb}{\partial y^2},\\
&\pp{c}{t}=\frac{a}{\lb(\tau)}\pp{\cb}{\tau}-\frac{ay\lb'(\tau)}{\lb(\tau)^2}\pp{\cb}{y}.
\end{aligned} \right. \]
To simplify notation we note that
\[ 1-y\frac{\lb'}{\lb} = 1 - y\frac{l'}{a} = 1-y\frac{\rg}{a}{\big(\ccb-\cinf\big)} \]
and define the functions
\begin{align*}
&\alpha(\ccb,y) := 1-y\frac{\rg}{a}{\big(\ccb-\cinf\big)}\quad\text{and}\\
&\beta(\ccb,\lb) := \frac{(a-g\lc)\lb\ccb - \lb(\rg\ccb+\rgtilde\lc)(\ccb-\cinf)}{a\lc}.
\end{align*}
Thus, after both space and time scaling of \eqref{eq:model} we get the dynamic
system
\begin{subequations}\label{eq:modely}
\begin{empheq}[left = \empheqlbrace\,]{align}
&\pp{\cb}{\tau} + \alpha(\ccb,y)\pp{\cb}{y} - \frac{D}{a}\frac{1}{\lb}\frac{\partial^2\cb}{\partial y^2} = -\frac{g}{a}\lb\cb, \label{eq:modely1}\\
&\dd{\ccb}{\tau} = \beta(\ccb,\lb) - \frac{D}{a\lc}\cb_y^-,\label{eq:modely2}\\
&\dd{\lb}{\tau}=\frac{\rg}{a}\lb(\ccb-\cinf),\label{eq:modely3}\\
&\frac{\mathrm{d}\tb}{\mathrm{d}\tau} = \frac{1}{a}\lb,\label{eq:modely4}\\
&\cb(0,\tau)=\cs(\tb(\tau)),\label{eq:modely5}\\
&\cb(1,\tau)=\ccb(\tau),\label{eq:modely6}\\
&\cb(y,0) =c^0(yl^0),\label{eq:modely7}\\
&\ccb(0)= c^0(l^0),\label{eq:modely8}\\
&\lb(0)=l^0,\label{eq:modely9}\\
&\tb(0) = 0.\label{eq:modely10}
\end{empheq}
\end{subequations}
which is defined for $y\in (0,1)$ and $\tau > 0$.

The positive effect of the time scaling for short axon lengths can for example be seen by comparing \eqref{eq:modelyonly1} with \eqref{eq:modely1}. In the latter there is one less reciprocal of $l$ in the advection and diffusion coefficients. Thus, the evolution of the system when the axon length $l$ is small is easier to resolve in the scaled time $\tau$ compared to the original time $t$.

Note that the PDE \eqref{eq:modely1} is linear in $\cb$. However, the coefficients in \eqref{eq:modely1} and the boundary conditions \eqref{eq:modely5}--\eqref{eq:modely6} depend on $\ccb$, $\lb$, and $\tb$ which are determined by the nonlinear ODEs \eqref{eq:modely2}--\eqref{eq:modely4}. Since, additionally the ODE \eqref{eq:modely2} depends on $\cb_y^-$, the model \eqref{eq:modely} defines a fully coupled nonlinear system. The same observations can be made for the system \eqref{eq:modelyonly}. In Sec.~\ref{sec:numerical_methods} we will apply a numerical method that decouples the approximations of the differential equations such that the aforementioned linearities can be utilized.

{Finally, we comment on an alternative time scaling where \eqref{eq:gamma} is replaced by
\begin{equation}
\Gamma_D(t):=D\int\limits_{0}^{t}\frac{\mathrm{d}s}{l(s)^2},
\label{eq:scaling_diffusion}
\end{equation}
which means that the former scaling with respect to advection velocity $a$ is replaced by a scaling with respect to the diffusion $D$. This gives instead of \eqref{eq:modely1} the PDE
\begin{equation}
\pp{\cb}{\tau} + \frac{a}{D}\alpha(\ccb,y) \lb\ \pp{\cb}{y} - \frac{\partial^2\cb}{\partial y^2} = -\frac{g}{D}\lb^2\cb.
\label{eq:scaling_diffusion_PDE}
\end{equation}
The corresponding ODEs are similarly given by multiplying the right-hand sides of \eqref{eq:modely2}--\eqref{eq:modely4} by $a\lb/D$. The function $\Gamma_D$ implements a more aggressive scaling promoting high resolution when the axon is short at the expense of poorer resolution during the time periods when $l$ is large. See also the remarks at the end of Sec.~\ref{sec:time_transformation_simulations}.}

\section{Numerical methods} \label{sec:numerical_methods}

We approximate the fully scaled system \eqref{eq:modely} using the method of
lines (MOL). To this end, we perform a spatial discretization in
Sec.~\ref{sec:semi-discretization}, which is followed by temporal
discretizations in Sec.~\ref{sec:full_discretization}. Note that the same
discretizations may be applied to system \eqref{eq:modelyonly} which is only
scaled in space. However, in Sec.~\ref{sec:time_transformation_simulations} we
will see that by using time scaling we largely gain in efficiency and
reliability. See also \citet{SDJTB1} for an explicit Euler discretization of
\eqref{eq:modelyonly}.

\subsection{Spatial discretization}\label{sec:semi-discretization}

The spatial interval $[0,1]$ is divided
into $M$ subintervals of size $\Delta y:=1/M$. The grid points are located at
$y_j:=j\Delta y$, $j=0,\ldots,M$. In particular, $y_M=1$ holds. To each grid
point we associate a concentration value $\ct_j = \ct_j(\tau) \approx
\cb(y_j,\tau)$.
We approximate the spatial derivatives of $\cb$ in the PDE \eqref{eq:modely1}
by second-order central finite differences, i.e., for $j = 1,\dots,M-1$,
\begin{equation}
\begin{aligned}
&\pp{\cb}{y}(y_j,\cdot) \approx \frac{\ct_{j+1} - \ct_{j-1}}{2\Delta y} \quad \text{and} \\
&\frac{\partial^2\cb}{\partial y^2}(y_j,\cdot) \approx \frac{\ct_{j+1} - 2\ct_{j} + \ct_{j-1}}{(\Delta y)^2}.
\label{eq:FD}
\end{aligned}
\end{equation}
Thus, the PDE \eqref{eq:modely1} is transformed into a system of $M-1$ MOL
ODEs:
\[ \begin{split}
\dd{\ct_j}{\tau} &= -\alpha(\ccb,y_j)\frac{\ct_{j+1} - \ct_{j-1}}{2\Delta y} \\
	&+ \frac{D}{a}\frac{1}{\lb}\frac{\ct_{j+1} - 2\ct_{j} + \ct_{j-1}}{(\Delta y)^2}-\frac{g}{a}\lb\ct_j, \quad j = 1,\dots, M-1.
\end{split} \]
Here $\ct_0(\tau)$ and $\ct_M(\tau)$ should be interpreted as $\cs(\tb(\tau))$ and $\ccb(\tau)$, respectively, according to the boundary conditions \eqref{eq:modely5}--\eqref{eq:modely6}. In the cone concentration ODE~\eqref{eq:modely2} we use a one-sided second-order approximation together with the continuity boundary condition \eqref{eq:modely6} to get:
\begin{equation}
\cb_y^- \approx \frac{3\ct_M - 4\ct_{M-1} + \ct_{M-2}}{2\Delta y} = \frac{3\ccb - 4\ct_{M-1} + \ct_{M-2}}{2\Delta y}.
\label{eq:FD_end_point}
\end{equation}
The above efforts combined replace the model~\eqref{eq:modely} by a system of
$M+2$ ODEs. This MOL discretization can be written on matrix form as
\begin{subequations} \label{eq:split}
\begin{empheq}[left = \empheqlbrace\,]{align}
\dd{\bct}{\tau} &= \bA(\bub,\by)\bct + \bB(\bub), \label{eq:split_1} \\
\dd{\bub}{\tau} &= \bF(\bct,\bub),   \label{eq:split_2}
\end{empheq}
\end{subequations}
with initial values given by \eqref{eq:modely7}--\eqref{eq:modely10} and a sampling of $c^0$: $\ct_j(0) = c^0(y_jl^0)$ for $j = 1,\dots,M-1$. We explain the notation of \eqref{eq:split} in what follows. To this end, introduce the vectors
\[\begin{aligned}
&\by := \bp y_1 & y_2 & \cdots & y_{M-1} \ep\trans, \\
&\bct(\tau) := \bp \ct_1(\tau) & \ct_2(\tau) & \cdots & \ct_{M-1}(\tau) \ep\trans, \\
&\bub(\tau) := \begin{pmatrix} \ccb(\tau)~~ & \lb(\tau)~~ & \tb(\tau) \end{pmatrix}^{\mathrm{T}}.
\end{aligned}\]
and define the tridiagonal matrix
\[ \bA :=
\bp a_{1,1}	\phantom{hh}	&		a_{1,2}		&		0			&		\cdots		&		0 \\
a_{2,1}	\phantom{hh}	&		a_{2,2}		&		a_{2,3}			&		\ddots		&		\vdots \\
0	\phantom{hh}	&		\ddots		&		\ddots		&		\ddots		&			0  \\
\vdots	\phantom{hh}	&		\ddots	&		a_{M-2,M-3}		&		a_{M-2,M-2}		&		a_{M-2,M-1} \\
0	\phantom{hh}	&		\cdots		&		0		&		a_{M-1,M-2}		&		a_{M-1,M-1} \ep\]
of size $(M-1)\times(M-1)$. The entries of $\bA = \bA(\bub,\by)$ depend on the solutions of the ODEs \eqref{eq:modely2}--\eqref{eq:modely4}. These entries are given by
\[ \begin{aligned}
&a_{j,j-1}(\bub,\by) := \frac{\alpha(\ccb,y_j)}{2\Delta y} + \frac{D}{a(\Delta y)^2}\frac{1}{\lb}, && j = 2, \dots, M-1,\\
&a_{j,j}(\bub,\by) := -\frac{2D}{a(\Delta y)^2}\frac{1}{\lb} - \frac{g}{a}\lb, && j = 1, \dots, M-1, \\
&a_{j,j+1}(\bub,\by) := -\frac{\alpha(\ccb,y_j)}{2\Delta y} + \frac{D}{a(\Delta y)^2}\frac{1}{\lb}, && j = 1, \dots, M-2.
\end{aligned} \]
Note how the sub and super diagonals vary with $y_j$: on row $j$ we input $y_j$
in $\alpha$ to get the correct matrix elements. Also note that on the first and
last row there are only two non-zero elements. Further, define the solution-dependent vector
\begingroup
\renewcommand*{\arraystretch}{1.0}
\[ \bB(\bub) := \bp \left( \frac{1}{2\Delta y}\alpha(\ccb,y_1) + \frac{D}{a(\Delta y)^2}\frac{1}{\lb} \right) \cs(\tb) \\
0 \\ \vdots \\ 0 \\ \left( -\frac{1}{2\Delta y}\alpha(\ccb,y_{M-1}) + \frac{D}{a(\Delta y)^2}\frac{1}{\lb} \right) \ccb \ep, \]
\endgroup
which is of size $M-1$ and contains the boundary conditions \eqref{eq:modely5}--\eqref{eq:modely6}. Finally, define the vector
\begingroup
\renewcommand*{\arraystretch}{1.3}
\[ \bF(\bct,\bub) := \begin{pmatrix}
\beta(\ccb,\lb) - \frac{D}{a\lc} \frac{3\ccb - 4\ct_{M-1} + \ct_{M-2}}{2\Delta y} \\
\frac{\rg}{a}\lb(\ccb-\cinf) \\
\frac1a \lb
\end{pmatrix} \]
\endgroup
corresponding to the right-hand sides of the ODEs
\eqref{eq:modely2}--\eqref{eq:modely4}. Thus, we arrive at the MOL
discretization \eqref{eq:split}, given by applying second-order finite
differences to the system \eqref{eq:modely}.

\subsection{Full discretizations}\label{sec:full_discretization}

Based on the semi-discretization \eqref{eq:split} we use the explicit Euler and
Peaceman-Rachford methods to construct two different full discretizations.
Denote the time step by $\Delta\tau$ and let $\tau^n:=n\Delta\tau$,
$n=0,1,\ldots, N$, where $N$ is the number of time steps used. At time
$\tau=\tau^n$, the concentration within the axon is approximated by the
numerically computed values $C_j^n \approx \ct_j(\tau^n) \approx
\cb(y_j,\tau^n)$, $j=1,\ldots,M-1$. The approximate growth-cone concentration
is denoted by $\Cc^n\approx\ccb(\tau^n)$, the approximate axon length by
$L^n\approx \lb(\tau^n)$, and the approximate (original) time by $t^n \approx
\tb(\tau^n)$. We gather these values in the vectors
\begin{align*}
\bC^n &:= \begin{pmatrix} C_1^n~~ & C_2^n & \hdots & C_{M-1}^n \end{pmatrix}^{\mathrm{T}}, \\
\bU^n &:= \begin{pmatrix} \Cc^n~~ & L^n~~ & t^n \end{pmatrix}^{\mathrm{T}}.
\end{align*}

By applying the explicit Euler temporal discretization to the spatial semi-discretization \eqref{eq:split} we get, for $n = 0,1,\dots, N-1$,
\begin{subequations} \label{eq:EE}
\begin{align}
\bC^{n+1} &= \bC^n + \Delta\tau\ \big(\bA(\bU^n,\by)\bC^n + \bB(\bU^n) \big), \label{eq:EE_PDE}\\
\bU^{n+1} &= \bU^n + \Delta\tau\ \bF(\bC^n,\bU^n).  \label{eq:EE_ODE}
\end{align}
\end{subequations}
Equation~\eqref{eq:EE} defines a time-marching scheme with initial values
\begin{subequations} \label{eq:IC}
\begin{align}
&C_j^0 := c^0(y_jl^0), \quad j=1,\ldots,M-1, \\
&\bU^0 := \bp c^0(l^0)~~ & l^0~~ & 0 \ep\trans.
\end{align}
\end{subequations}

In the original coordinates, we have $\cc(t^n) \approx \Cc^n$ and $l(t^n) \approx L^n$ at the time points $t^n$ and for the concentration distribution in the axon
$c(x_j^n,t^n) \approx C_j^n$ where $x_j^n := y_jL^n = j\Delta yL^n$.

The explicit Euler method \eqref{eq:EE} defines computations that are simple to
implement since only old values of the unknowns are used. However, a problem
with this method (and explicit methods in general) is that, given $\Delta
y=1/M$, the time step $\Delta\tau$ has to be chosen small to avoid numerical
instabilities. When diffusion is present these time step restrictions are very
prohibitive and result in large CPU times. In the classical analysis of the
explicit Euler method applied to diffusion--advection--reaction equations a so
called CFL condition must be fulfilled to have stability. Here we have an
additional complication due to the coupling to the ODEs
\eqref{eq:modely2}--\eqref{eq:modely4}.
Assume that the scheme \eqref{eq:EE} produces a numerical solution that satisfies
\begin{align}\label{eq:Lbounded}
&L^n \geq \lmin > 0 \quad \text{for } n = 0,1,\dots, N, 
\end{align}
where $\lmin$ is a constant. Then the explicit Euler method \eqref{eq:EE} has
the CFL stability criterion
\begin{equation}
\Delta \tau \leq \frac{a}{2D}\lmin(\Delta y)^2.
\label{eq:CFL}
\end{equation}

To summarize, while the advantage of the explicit scheme \eqref{eq:EE} is its
simple implementation, the disadvantage is the long computation times required
due to the factor $(\Delta y)^2$ in the right-hand side of \eqref{eq:CFL}. With
$\lmin=\SI{1}{\micro\metre}$ and the parameter values of Table~\ref{tab:1}, the
CFL condition \eqref{eq:CFL} is for $\Delta y=1/M=1/100$ given by $\Delta\tau
\leq \num{5e-8}$. This should be compared with the long simulation times
usually needed to get close to steady state. See for example the simulation
presented in Fig.~\ref{fig:solution} with end time $t^N \geq \SI{6e8}{s}
\approx 19$ years and where the axon length is at its longest $L^n \approx
\SI{0.08}{\metre}$. Using \eqref{eq:modely4} we can deduce that $\Delta t$ is,
at its largest, approximately \SI{0.4}{s}. This means that we need more than
$15\cdot 10^8$ time steps just to ensure the stability of explicit Euler.
Additionally, if we want higher accuracy in space (as in the
aforementioned simulation) the number of time steps $N$ needed grows
quadratically with $M$ at the same time as there are $M$ function computations
at each time point. In other words, with explicit Euler
time stepping, halving $\Delta y$, means that the CPU time increases with a
factor approximately $2^3=8$.

To avoid the aforementioned problems we can instead use an unconditionally
stable implicit scheme, like the implicit Euler method. Such a method needs no
stability restriction on $\Delta \tau$. However, for an implicit method the
semi-discretization~\eqref{eq:split_1} defines a large system of equations
which is coupled with the nonlinear ODEs of \eqref{eq:split_2}. This means that
in every time step a nonlinear equation solver needs to be applied to the full
system.

We propose to instead use the Peaceman-Rachford splitting method for
time discretization, cf.\ \citet{Douglas1955,Peaceman1955,Hundsdorfer,Hansen2013}. This method needs neither a stability constraint on
$\Delta \tau$ (as a function of $\Delta y$), nor the numerical solution of a
large nonlinear system at each time step. Furthermore, while the explicit and
implicit Euler methods are first-order accurate, the Peaceman-Rachford method
is second-order accurate. Finally, in contrast to many other splitting methods, the Peaceman--Rachford scheme preserves the steady states of the system that it approximates, cf.\ \citet[Sec.~IV.3.1]{Hundsdorfer}. This preservation property is of utmost importance for our investigations in Sec.~\ref{sec:parameter_studies}.

Taking one time step of size $\Delta \tau$ using the
Peaceman-Rachford splitting scheme consists of, in sequence, solving for
$\bU^{n+1/2}$, $\bC^{n+1/2}$, $\bC^{n+1}$, and $\bU^{n+1}$, respectively, in
the following equations:
\begin{subequations} \label{eq:PR}
\begin{align}
&\bU^{n+\frac12} = \bU^n + \frac{\Delta\tau}{2} \bF(\bC^n,\bU^n),  \label{eq:PR_ODE_EE} \\
&\bC^{n+\frac12} = \bC^n + \frac{\Delta\tau}{2}\!\! \left( \bA(\bU^{n+\frac12},\by)\bC^{n+\frac12} + \bB(\bU^{n+\frac12}) \right), \label{eq:PR_PDE_IE}\\
&\bC^{n+1} = \bC^{n+\frac12} + \frac{\Delta\tau}{2}\!\! \left( \bA(\bU^{n+\frac12},\by)\bC^{n+\frac12} + \bB(\bU^{n+\frac12}) \right), \label{eq:PR_PDE_EE}\\
&\bU^{n+1} = \bU^{n+\frac12} + \frac{\Delta\tau}{2} \bF(\bC^{n+1},\bU^{n+1}).  \label{eq:PR_ODE_IE}
\end{align}
\end{subequations}
Here, $\bC^n$ and $\bU^n$ are known from the previous time step or, for $n = 0$, from the initial conditions \eqref{eq:IC}.

Some comments about the Peaceman--Rachford method are appropriate. First note that the updates \eqref{eq:PR_ODE_EE} and \eqref{eq:PR_PDE_EE} are explicit Euler steps similar to \eqref{eq:EE_PDE} and \eqref{eq:EE_ODE} and they are therefore cheap to compute. The other two updates, \eqref{eq:PR_PDE_IE} and \eqref{eq:PR_ODE_IE}, are implicit Euler steps. However, they are implicit only in $\bC^{n+1/2}$ and $\bU^{n+1}$, respectively. As a consequence, performing the update \eqref{eq:PR_PDE_IE} only amounts to solving the linear system of equations
\begin{equation}
\left( \bI - \frac{\Delta\tau}{2} \bA(\bU^{n+\frac12},\by) \right) \bC^{n+\frac12} = \bC^n + \frac{\Delta\tau}{2} \bB(\bU^{n+\frac12})
\label{eq:PR_PDE_IE_linear_system}
\end{equation}
for $\bC^{n+1/2}$, where $\bI$ is the $(M-1)\times (M-1)$ identity matrix. This
should be compared with a fully implicit method, for which the corresponding
system to be solved in each time step is nonlinear (and also slightly larger,
consisting of $M+2$ equations). Finally, performing the update
\eqref{eq:PR_ODE_IE} only means solving a system of three nonlinear equations.
For that, a standard nonlinear equation solver, like Newton's method, can be
applied for rapid solution. Such a solver requires a tolerance
parameter to determine how many local iterations are needed. However, since the nonlinear equation system \eqref{eq:PR_ODE_IE} is tiny compared to the linear system \eqref{eq:PR_PDE_IE} the solution of the former can be done in a negligible time. Thus, optimizing the tolerance parameter is superfluous; we may choose it sharp without significantly affecting the overall efficiency of the scheme.

For the update \eqref{eq:PR_PDE_IE} to be well-defined the matrix $\bI - \Delta\tau/2 \cdot \bA(\bU^{n+1/2},\by)$ must be invertible, we give a sufficient condition in the following lemma.
{\begin{lemma} \label{lem:A_negative_definite}
Assume that the scheme \eqref{eq:PR} produces a numerical solution that satisfies $L^{n+1/2} > 0$ and
\begin{equation} 
|\Cc^{n+\frac12} - \cinf| \leq \gamma, \quad \text{for } n = 0, 1, \dots, N,
\label{eq:lemma_1_solution_condition} \end{equation}
and for some positive constant $\gamma$. Then, if
\begin{equation}
\Delta\tau < \frac{4a}{\rg} \cdot \frac{1}{\gamma},
\label{eq:lemma_1_step_condition}
\end{equation}
the linear system of equations \eqref{eq:PR_PDE_IE_linear_system} has a unique solution at each time step $n$.
\end{lemma}}
See Appendix~\ref{app:Lemma_A_negative_definite} for a proof. Note that the condition \eqref{eq:lemma_1_step_condition} is well behaved in the sense that it is not affected by small values of $\Delta y$. In fact, consider the simulations performed in Sec.~\ref{sec:dynamical_properties}, the solution fulfils the bound \eqref{eq:lemma_1_solution_condition} with $\gamma = \SI{11.9e-3}{\mol\per\cubic\metre}$. With this constant and with the parameter values as in Table~\ref{tab:1} the time step condition \eqref{eq:lemma_1_step_condition} reads {$\Delta \tau < 0.18$. We emphasize that using this bound on $\Delta\tau$ does not guarantee that the entire numerical scheme works, e.g.\ the implicit Euler step \eqref{eq:PR_ODE_IE} for the ODEs may require smaller values on $\Delta\tau$. However, further investigations, cf.\ Fig.~\ref{fig:convergence}, yield that much smaller time steps are needed for accuracy and mean no severe restriction in CPU time.}

{Note that if the scaling with diffusion $D$ \eqref{eq:scaling_diffusion} is used instead of the one with advection $a$ \eqref{eq:gamma}, the lemma should be changed in the following way: We have to assume that there is a constant $l^\text{max}$ such that $L^{n+1/2} \leq l^\text{max}$ for all $n$, and then the condition \eqref{eq:lemma_1_step_condition} is replaced by $\Delta \tau < 4D/(\rg l^\text{max}\gamma)$.}

\section{Simulations} \label{sec:simulations}
In this section we present numerical simulations performed with the methods in
Sec.~\ref{sec:numerical_methods}. The purpose is twofold: Firstly, in
Secs~\ref{sec:dynamical_properties}--\ref{sec:time_transformation_simulations}
we examine the dynamics of the system \eqref{eq:model} and how it affects the
choice of numerical method.
Secondly, in Sec.~\ref{sec:parameter_studies}, we use the efficient Peaceman--Rachford discretization to perform parameters studies. That is, we vary the parameter values of Table~\ref{tab:1} to investigate the sensitivity of the dynamical solution with respect to each parameter.

All results plotted in this section are numerical approximations, however, for
the sake of brevity we will not use the notation of
Sec.~\ref{sec:numerical_methods} but rather refer to each approximation via the
continuous variable that is approximated. For example, in
Fig.~\ref{fig:solution_l} a Peaceman--Rachford approximation $L^n$, $n = 0, 1,
\dots, N$ of $l$ given by \eqref{eq:PR} is plotted but the approximation is
referred to by $l$.

{The numerical schemes presented in Sec.~\ref{sec:numerical_methods} and used in the current section have been implemented in MATLAB (R2014a). The code is available from the ModelDB database with accession number 187687\footnote{URL: \url{https://senselab.med.yale.edu/ModelDB/showModel.cshtml?model=187687}}.}

\subsection{Biological, physical, and numerical constants} \label{sec:settings}
In our investigations, biological and physical as well as numerical parameters
will be varied depending on the inquiry at hand. The nominal parameter values
and initial and boundary conditions given here and in Table~\ref{tab:1} are
used except when something else is explicitly stated. Further, when nothing
else is stated, we use the Peaceman--Rachford discretization \eqref{eq:PR} to
approximate the spatial semi-discretization \eqref{eq:split} of the fully
scaled system \eqref{eq:modely}. Among the methods presented in this article
this is by far the most efficient and accurate approximation of system
\eqref{eq:model}, as we shall see in Sec.~\ref{sec:efficiency_of_PR} and
Sec.~\ref{sec:time_transformation_simulations}.

In all performed simulations the time-dependent soma concentration $\cs(t)$ is
chosen piecewise constant as
\begin{equation}\label{eq:cs}
 \cs(t) := \left\{ \begin{aligned}
&2 \cinf = \SI{23.80}{\milli\mol\per\cubic\metre}, &&0 \leq t < \SI{2e8}{\second}, \\
&\frac{\cinf}{2} = \SI{5.95}{\milli\mol\per\cubic\metre}, &&\SI{2e8}{\second} \leq t < \SI{4e8}{\second}, \\
&2 \cinf = \SI{23.80}{\milli\mol\per\cubic\metre}, &&\SI{4e8}{\second} \leq t.
\end{aligned} \right.
\end{equation}
With this choice of $\cs$ we will observe both axon expansion and contraction.
Additionally, note that we initially have $\cs(t)/\cinf = 2$ which means that
the axon will grow regardless of the initial length $l^0$, cf.\ \citet[Thm~4.1
and Fig.~9]{SDJTB1}.

Since we are interested in the growth of the axon from a small length to its
steady state (and possible contraction due to a decrease in the soma
concentration), we choose a small initial length $l^0 = \SI{1}{\micro\metre}$.
For such small axon lengths it seems reasonable that the initial tubulin
concentration $c^0$ along the axon is constant and equal to the initial soma
concentration. Therefore we make the simple choice $c^0 = c^0(x) =
\SI{23.80e-3}{\mol\per\cubic\metre}$ for the initial concentration profile.

In each of our investigations we specify an end time $T$. The simulations will be performed until the first $n$ such that $t^n \geq T$ and we denote this value of $n$ by $N$. Due to the adaptivity in time caused by the time scaling, which depends on the solution, we cannot expect that $t^N$ is equal to $T$. However approximations of $c(x,T), \cc(T)$, and $l(T)$ can be found by simple interpolations.

\subsection{Dynamical properties of the model} \label{sec:dynamical_properties}

In Sec.~\ref{sec:model_transformation} we stated that system~\eqref{eq:model}
exhibits dynamical phenomena on different time scales. In this section we
verify this claim by showing numerical simulations on very fine grids as to
minimize the influence of numerical artefacts on the exact dynamics of the
systems. The numerical method, the biological parameters, and the initial data
are chosen as described in Sec.~\ref{sec:settings}. For the spatial
discretization we use $M = \num{e4}$ meaning $\Delta y = \num{e-4}$. (Recall
that there is no CFL condition for the Peaceman--Rachford scheme).

\begin{figure}[tb]
\centering
\subfigure{
		\topinset{\bfseries(a)}{\includegraphics[width=.95\columnwidth]{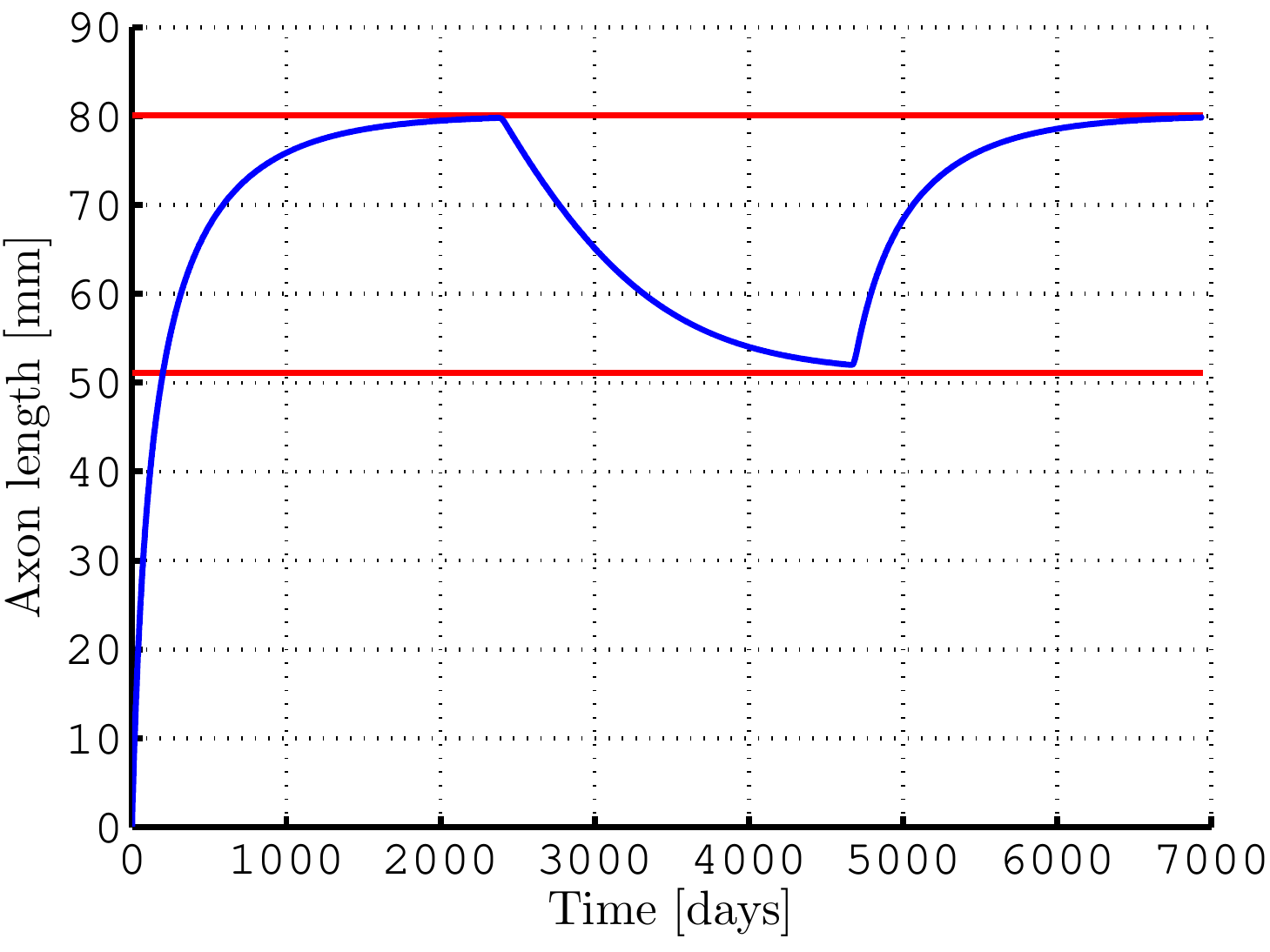}}{0mm}{0mm}
		\label{fig:solution_l}}
\subfigure{
		\topinset{\bfseries(b)}{\includegraphics[width=.95\columnwidth]{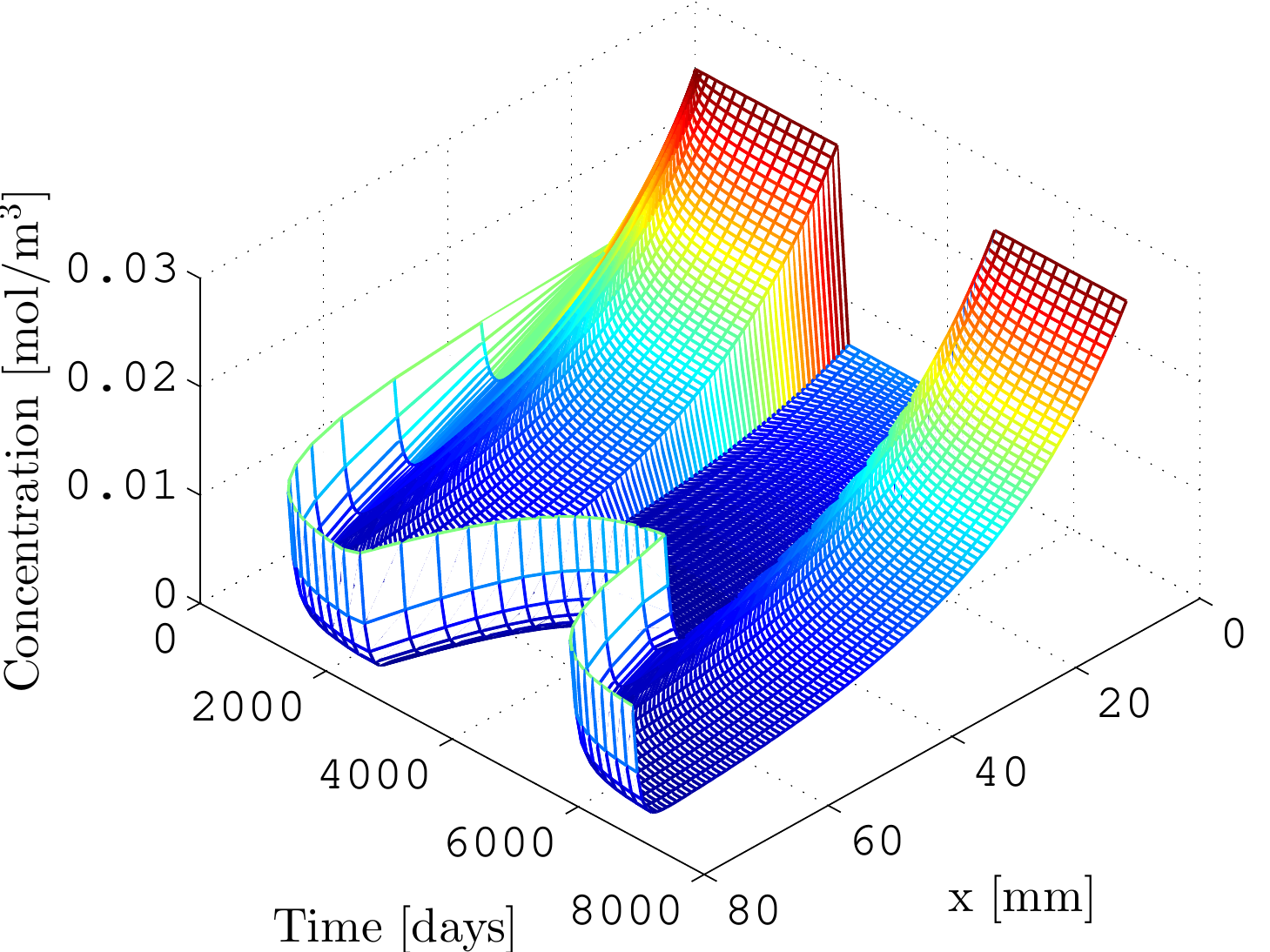}}{0mm}{0mm}
		\label{fig:solution_c}}
\caption{Axon growth and tubulin contraction as a result of a piecewise constant soma concentration $\cs(t)$, which can be seen in plot (b) at $x = 0$. In (a) the axon length $l(t)$ is plotted in blue and the red lines are the steady-state lengths corresponding to the two different values of soma concentration. In (b) the tubulin concentration $c(x,t)$ in the axon is plotted over time and space. Note the characteristic profile of the axon concentration, cf.\ \citet[Sec.~4]{SDJTB1}. Further, note how slowly the axon length responds to changes in $\cs(t)$ and how long time it takes for the axon to grow to its steady state. {See also Figs. \ref{fig:solution_c_slices_at_x_s} and \ref{fig:solution_c_slices_at_t_s} in Appendix B for two-dimensional slices of (b) at different values of $x$ and $t$.}}
\label{fig:solution}
\end{figure}
To investigate the behaviour on large time scales and the convergence to steady
state we choose the end time $T = \SI{6e8}{\second} \approx 19$~years and use
the small time step $\Delta \tau = \num{5e-4}$. The results are plotted in
Fig.~\ref{fig:solution} where we can observe how the variations in the soma
concentration $\cs$ cause axon expansion as well as contraction.

\begin{figure}[tb]
\centering

\subfigure{
		\topinset{\bfseries(a)}{\includegraphics[width=.95\columnwidth]{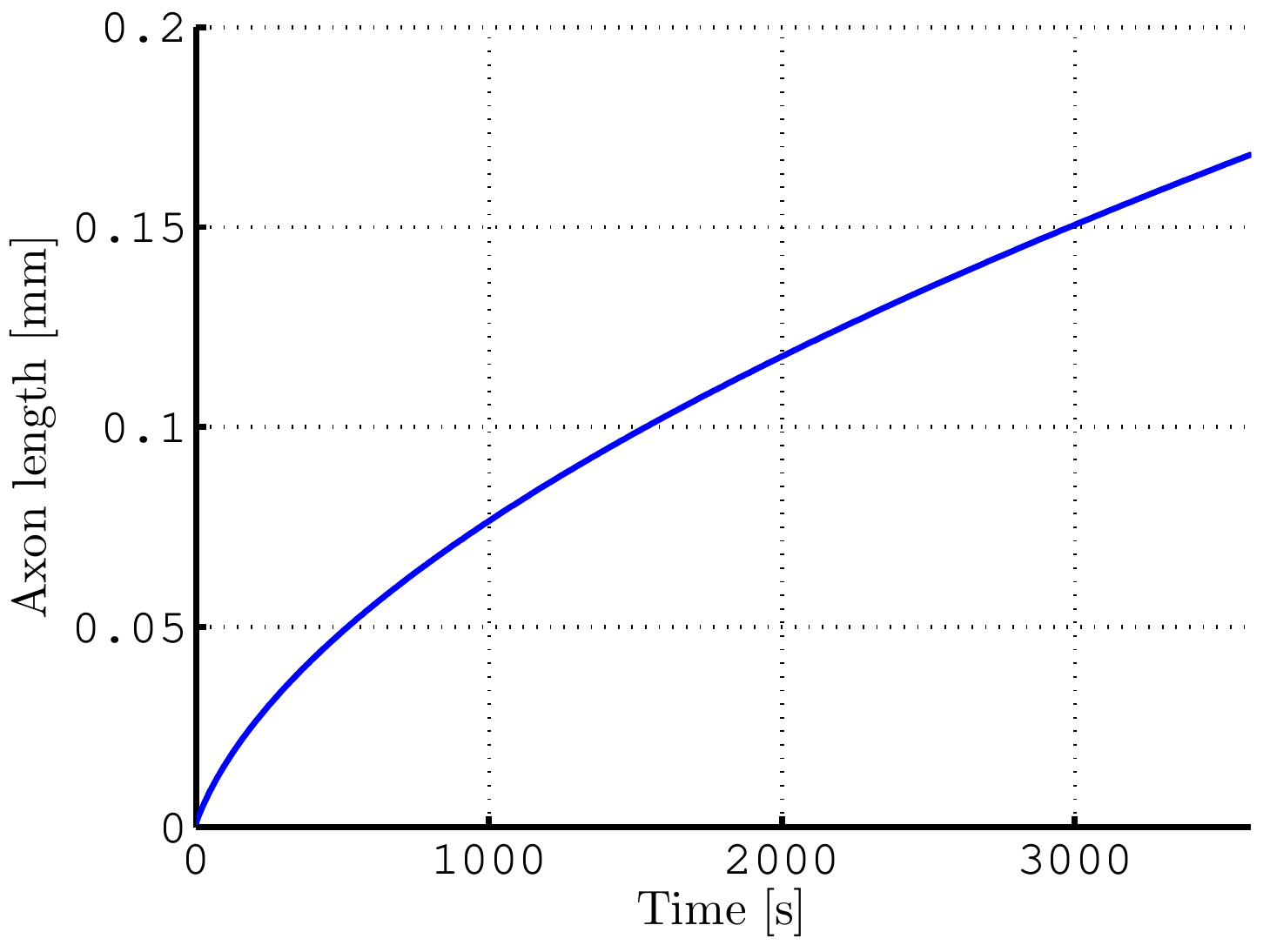}}{0mm}{0mm}
		\label{fig:solution_short_l}}
\subfigure{
		\topinset{\bfseries(b)}{\includegraphics[width=.95\columnwidth]{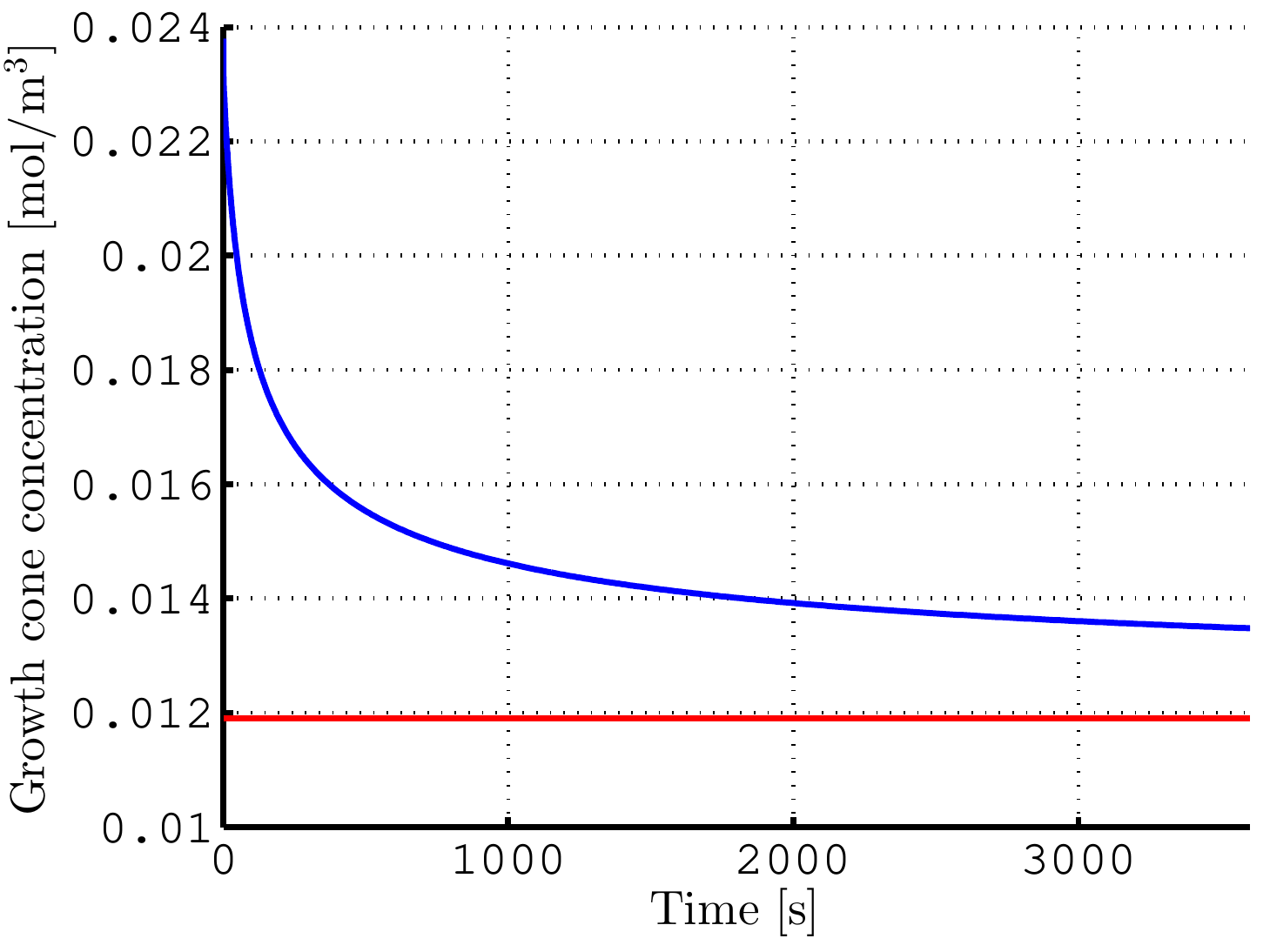}}{0mm}{0mm}
		\label{fig:solution_short_cc}}
\caption{(a) Axon length $l(t)$ and (b) cone concentration $\cc(t)$ during the first hour. In (a) we observe a fast growth of the axon to more than a hundred times its initial length. In (b) we see the fast convergence of the cone concentration $\cc(t)$ from $\cc(0) = 2\cinf = \SI{23.80e-3}{\mol\per\m^3}$ to its steady state $\cinf = \SI{11.90e-3}{\mol\per\m^3}$ (marked with a red line).}
\label{fig:solution_short}
\end{figure}
In Fig.~\ref{fig:solution_short} we can observe the fast transient behaviour of
system \eqref{eq:model} for small times. Here the end time is chosen as $T =
\SI{3600}{\second} = \SI{1}{\hour}$, tiny compared to the previous simulation.
Similarly, we here use a much smaller time step $\Delta \tau = \num{e-5}$ than
in Fig.~\ref{fig:solution}. Also note that this choice of $T$ gives
constant $\cs(t) = 2\cinf$. It is remarkable how close the cone concentration
$\cc$ is to its steady-state value $\cinf = \SI{11.90e-3}{\mol\per\m^3}$
already after \SI{1}{\hour}, whereas it takes more than a thousand days for the
axon length $l$ to come close to its steady state $\SI{80.10}{\milli\metre}$.

The evolution on these different time scales is one of the major problems for a numerical method to manage and it is an important motivation for the introduction of the time scaling \eqref{eq:ysofxt}. Even with time scaling, the fast transient of $\cc$ is the most difficult phenomenon for our numerical methods to resolve and therefore it is the main source of errors.

\subsection{Accuracy of the scaled Peaceman--Rachford scheme} \label{sec:efficiency_of_PR}
In this section we use numerical simulations to investigate the efficiency of the time discretization \eqref{eq:PR} as an approximation of model \eqref{eq:model}. We shine some light on why this approximation is more efficient than standard methods such as explicit Euler.

We first consider the end time $T = \SI{86400}{\second} = 1$~day so that we capture the transient of $\cc$. Note that this $T$ means that $\cs$ is constant and equal to $2\cinf = \SI{23.80e-3}{\mol\per\cubic\metre}$. Further, since we want to emphasize the error due to temporal discretization we choose a fine spatial grid $M = \num{e4}$, $\Delta y = \num{e-4}$. The errors are approximated by comparing the numerical solutions with a reference solution. The latter is given by using the same discretization scheme \eqref{eq:PR} on a very fine grid both in time, $\Delta\tau = \num{e-6}$, and in space, $\Delta y = \num{e-5}$. 

\begin{figure}[tb]
\centering\includegraphics[width=.95\columnwidth]{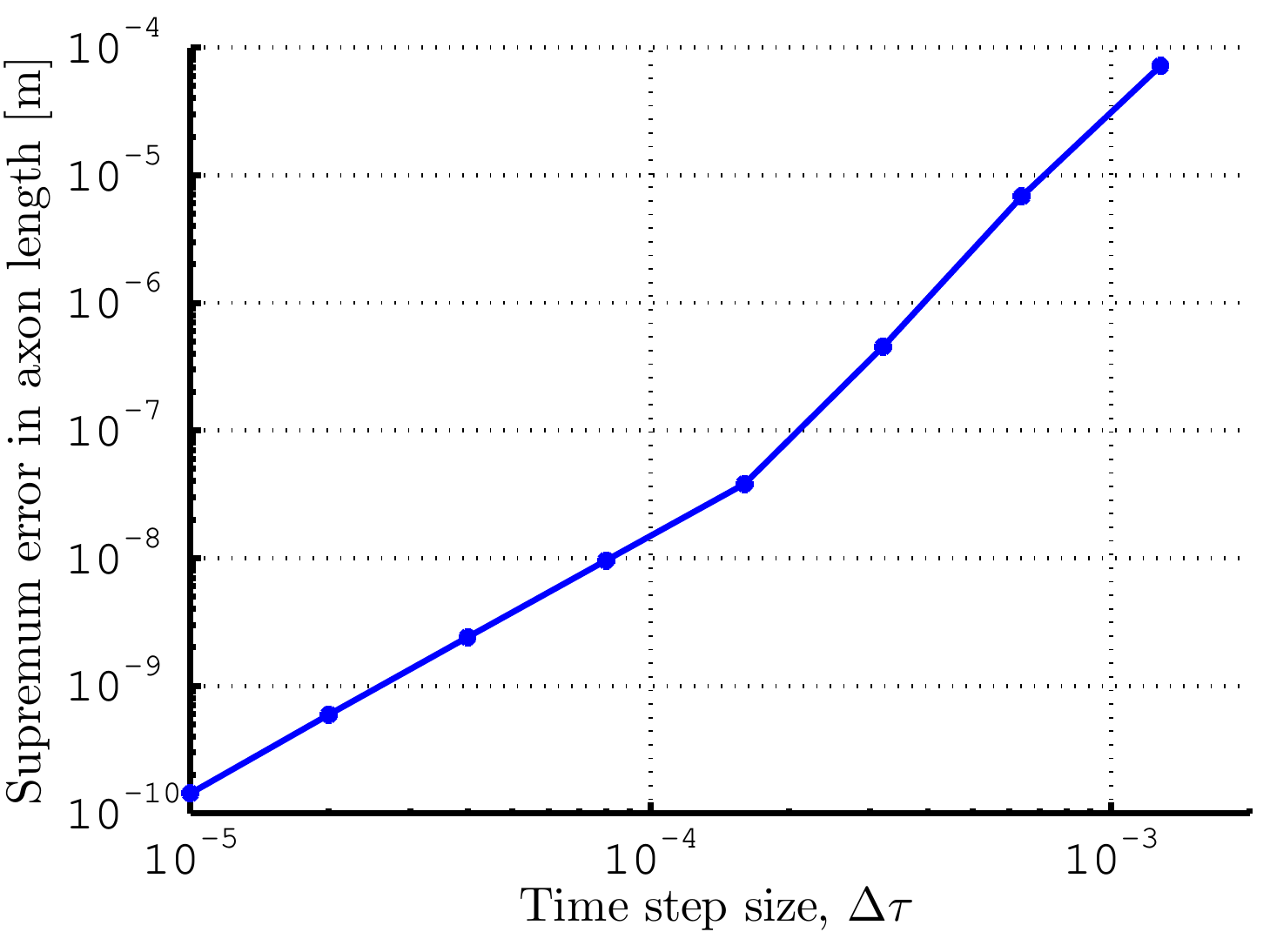}
\caption{Temporal convergence plot for the Peaceman--Rachford scheme.
Simulations are performed until the end time $T = \SI{86400}{\second} = 1$~day,
during which the axon growths to a length of approximately \SI{1.2}{\milli\metre}.
For each value of $\Delta \tau$, the maximal error (over time) in the axon length is plotted. We observe the expected second-order convergence. {For the leftmost point in the plot the CPU time is about five minutes and for the rightmost it is about three seconds (on a normal desktop computer).}}
\label{fig:convergence}
\end{figure}
In Fig.~\ref{fig:convergence} we see the results of a temporal convergence study for the Peaceman--Rachford discretization \eqref{eq:PR}. That is, we consider a range of different time step sizes, $\Delta \tau = 2^k\ee{-5},\ k = 0,\dots,7$, and perform simulations for each of them. Then, for each value of $\Delta \tau$ the error in axon length $l$ (given by taking the supremum norm over time) is plotted in Fig.~\ref{fig:convergence} over the corresponding $\Delta \tau$. Considering the slope of the curve when $\Delta \tau \lesssim \num{e-4}$ we observe the expected second-order convergence. The steeper slope for larger values of $\Delta \tau$ can be explained by the fact that the transient of $\cc$ is not properly resolved for those big time steps.

\subsection{The need for time transformation} \label{sec:time_transformation_simulations}

Recall the discussion in the beginning of Sec.~\ref{sec:numerical_methods},
where we concluded that we can approximate the spatially scaled system
\eqref{eq:modelyonly} with finite differences and Peaceman--Rachford using the
same procedure as when approximating the fully scaled system \eqref{eq:modely}
in Sec.~\ref{sec:numerical_methods}. This gives a numerical scheme without the
adaptivity in time granted by the time scaling \eqref{eq:gamma}. When the
simulation presented in Fig.~\ref{fig:convergence} is repeated with this
approximation no reasonable results are obtained due to the large time steps
taken during the first parts of the simulations. This shows the strength of
using the time scaling \eqref{eq:gamma} to get a finer temporal resolution for
small axon lengths.

\begin{figure}[tb]
\centering\includegraphics[width=.95\columnwidth]{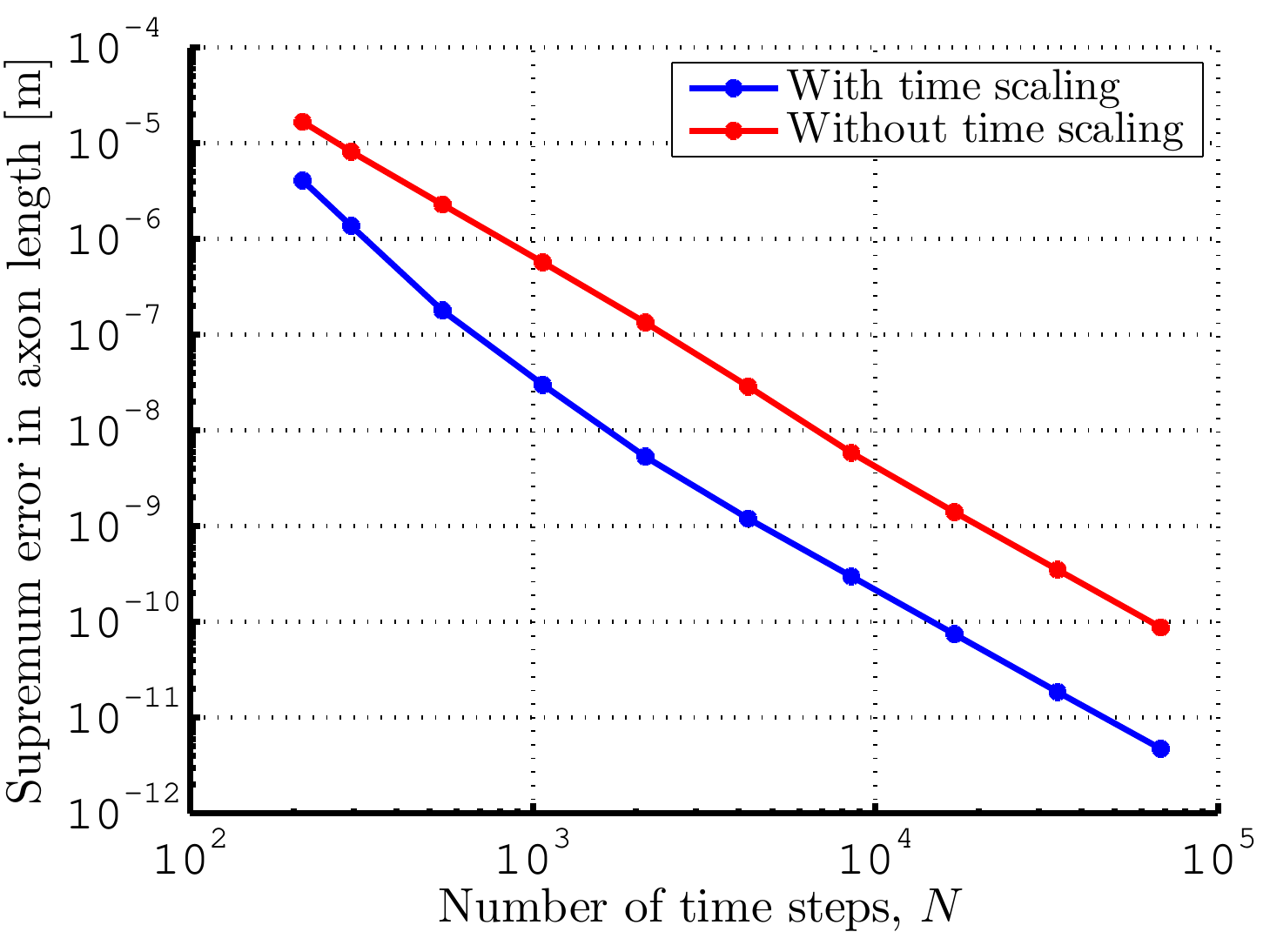}
\caption{Temporal convergence plots for the Peaceman-Rachford scheme applied to the
systems \eqref{eq:modelyonly} (red) and \eqref{eq:modely} (blue), respectively.
Simulations are performed with the small end time $T = \SI{60}{\second}$ giving
a final axon length of approximately \SI{11}{\micro\metre}. For each discretization
and each value of $\Delta \tau$ the maximal error (over time) in axon length is plotted.
With this fine resolution we can observe second-order convergence also when time scaling
is not used.}
\label{fig:convergence_60}
\end{figure}
When time scaling is not used, really small time steps are needed to properly
resolve the transient of the cone concentration $\cc$. In
Fig.~\ref{fig:convergence_60} results are plotted from simulations similar to
those presented in Fig.~\ref{fig:convergence}, but with a really small end time
$T = \SI{60}{\second}$. This $T$ is chosen to be able to complete simulations
with small enough time step sizes within reasonable CPU times. Also a finer
reference solution is needed: we construct it in the same way as described
above but here with $\Delta \tau = \num{e-7}$. To be able to compare the errors
given when time scaling is used with those given when it is not we plot the
errors over the number of time steps used, see Fig.~\ref{fig:convergence_60}.
For each $\Delta \tau = 2^k\ee{-6},\ k = 1,\dots,10$, we perform simulations
using time scaling and for each value of $\Delta \tau$ we store the number of
time steps used $N$. Then, for each $N$, a simulation is performed without time
scaling using $N$ time steps. We recall that in model \eqref{eq:modelyonly} $t$
is the independent time variable, which gives us $\Delta t = T/N$ for the
latter simulations.

The results in Fig.~\ref{fig:convergence_60} show that we can achieve
second-order convergence for the Peaceman--Rachford scheme even without using
time scaling. However, the size of the time steps needed are too small for this
discretization to be of any practical use. Therefore time scaling is preferable
and this is what we use for our parameter studies in
Sec.~\ref{sec:parameter_studies}.

The simulations presented in Fig.~\ref{fig:convergence_60} can be performed
using the explicit Euler method \eqref{eq:EE}. However, even when using time
scaling, we would need to choose $\Delta \tau < \num{5e-12}$ to fulfil
the CFL condition \eqref{eq:CFL}. With our implementation running on an
ordinary desktop computer, such a simulation would take about a quarter of a
year to perform whereas the Peaceman--Rachford simulation with $\Delta \tau =
1024\ee{-6}$ takes half a second and that with $\Delta \tau = 2\ee{-6}$ takes
about two minutes. Thus, explicit Euler is not an efficient choice, neither is
implicit Euler due to its expensive time steps, cf.\
Sec.~\ref{sec:full_discretization}. The same conclusions may be drawn for any
explicit method or implicit method that considers the whole axon growth model
at once.

\begin{figure}[tb]
\centering
\subfigure{
		\topinset{\bfseries(a)}{\includegraphics[width=.95\columnwidth]{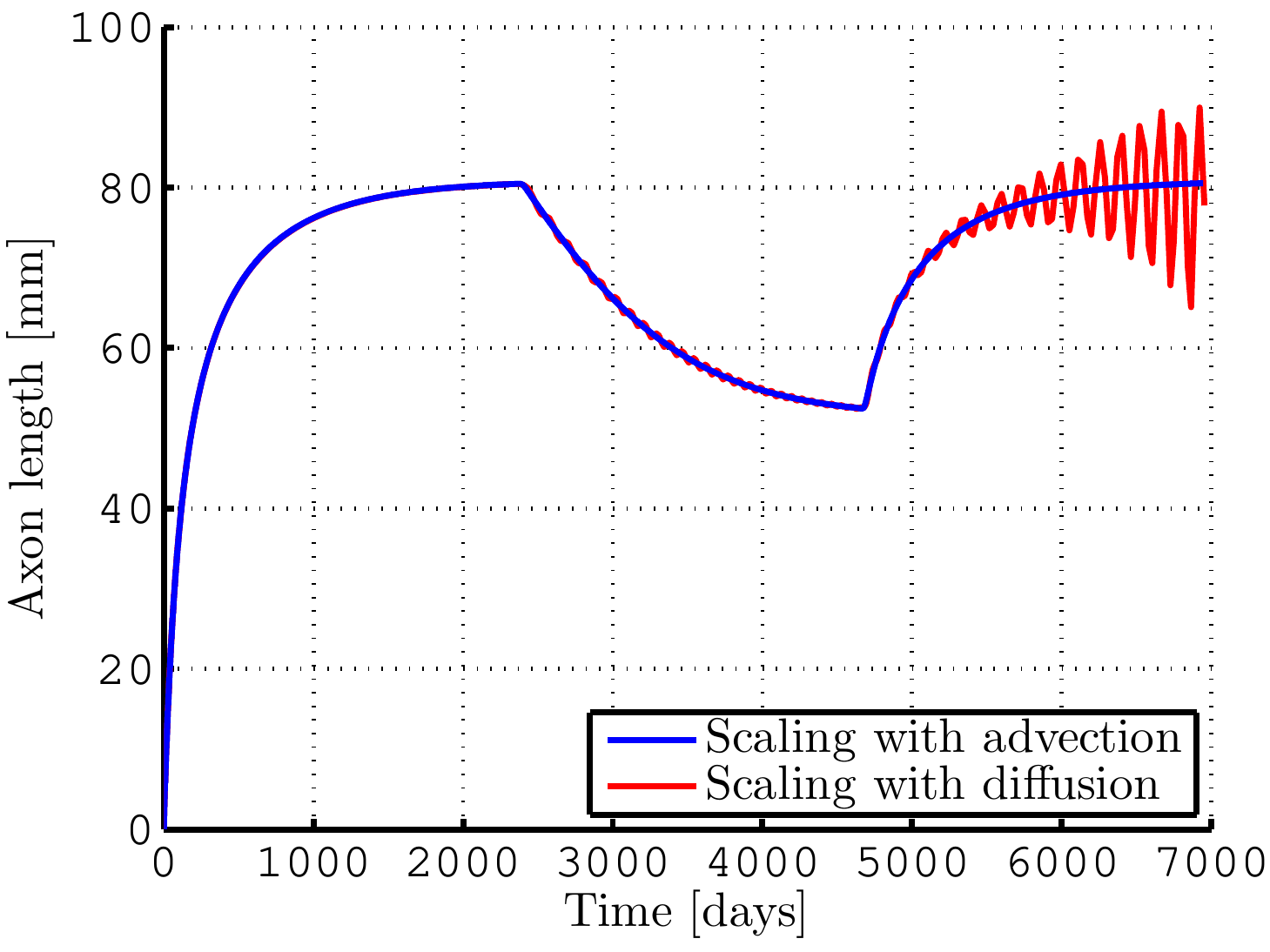}}{0mm}{0mm}
		\label{fig:testing_diffusion_scaling_l_not_zoomed}}
\subfigure{
		\topinset{\bfseries(b)}{\includegraphics[width=.95\columnwidth]{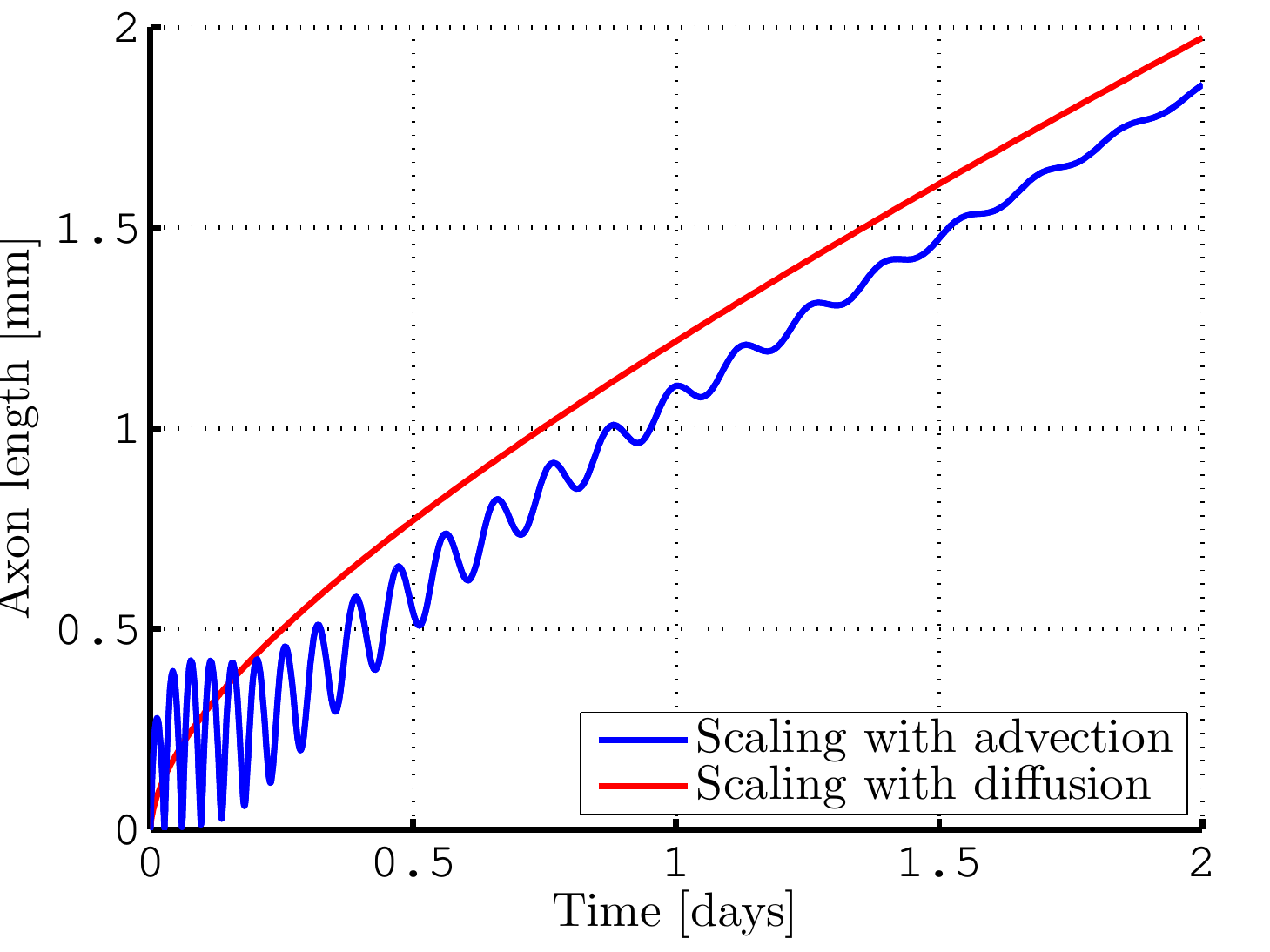}}{0mm}{0mm}
		\label{fig:testing_diffusion_scaling_l_zoomed}}
\caption{{Plot (a) shows that both scalings yield similar large-scale growth, but the diffusion scaling may develop errors for large axons. Zooming in at short times and lengths, plot (b) reveals that errors may arise for the advection scaling.}}
\label{fig:testing_diffusion_scaling_l}
\end{figure}
{Finally, we give a short comment on the effects of using time scaling with $D$ \eqref{eq:scaling_diffusion} instead of with $a$ \eqref{eq:gamma}. We test both scalings using the settings described in Sec.~\ref{sec:settings} with $\Delta y = 10^{-3}$ and $T = \SI{6e8}{\second} \approx 19$~years. For the advection scaling we choose $\Delta\tau = 6\ee{-3}$ and for the diffusion scaling $\Delta\tau = 4\ee{-3}$. These time steps are chosen large so that observable errors are produced. Furthermore, these choices mean that approximately the same number of time steps are used by both methods (approximately \num{16e3} steps), i.e.\ about the same amount of CPU time is used by each (a couple of seconds). The results are plotted in Fig.~\ref{fig:testing_diffusion_scaling_l} where we observe that the relative errors when using respective scaling are approximately the same. However, as one would expect the temporal location of the errors differ: The scaling with $a$ has its largest (relative) errors when the axon is short, whereas the scaling with $D$ has its largest errors when the axon is long. For the latter scaling we could also observe from our simulations that, after half the time steps were used, less than \SI{9}{\second} of simulated time $t$ had passed and the axon had only grown from \SI{1}{\micro\metre} to \SI{2.7}{\micro\metre}. As we are interested in the convergence to steady state we will continue to use the scaling with advection \eqref{eq:gamma} for our parameter studies in the upcoming section.
}

\subsection{Parameter studies} \label{sec:parameter_studies}

We shall now use the efficient and reliable Peaceman--Rachford
scheme~\eqref{eq:PR} to investigate the dependency of the axon growth dynamics
with respect to the parameters of the model. That is, we vary one, or a few, parameters at
a time performing simulations for a range of different values while keeping the
other parameters constant at their nominal values. The parameter study in this section
may be regarded as an extension of those performed in \citet[Sec.~5]{SDJTB1}
for steady-state solutions, and the interested reader may benefit from having
those results at hand. In that work, the steady states were presented for
different parameter values, here we consider the dynamical convergence to these
steady states. For easy comparison similar ranges of parameter values are used
whenever applicable. We focus on the axon length $l$ since the tubulin
concentrations tend fast to their steady-state appearances: the concentration
along the axon $c$, to its characteristic profile, cf.\
Fig.~\ref{fig:solution_c}, and the concentration in the growth cone, $\cc$, to
$\cinf$, cf.\ Fig.~\ref{fig:solution_short_cc}.

\begin{figure}[tb]
\centering
\subfigure{
		\topinset{\bfseries(a)}{\includegraphics[width=.95\columnwidth]{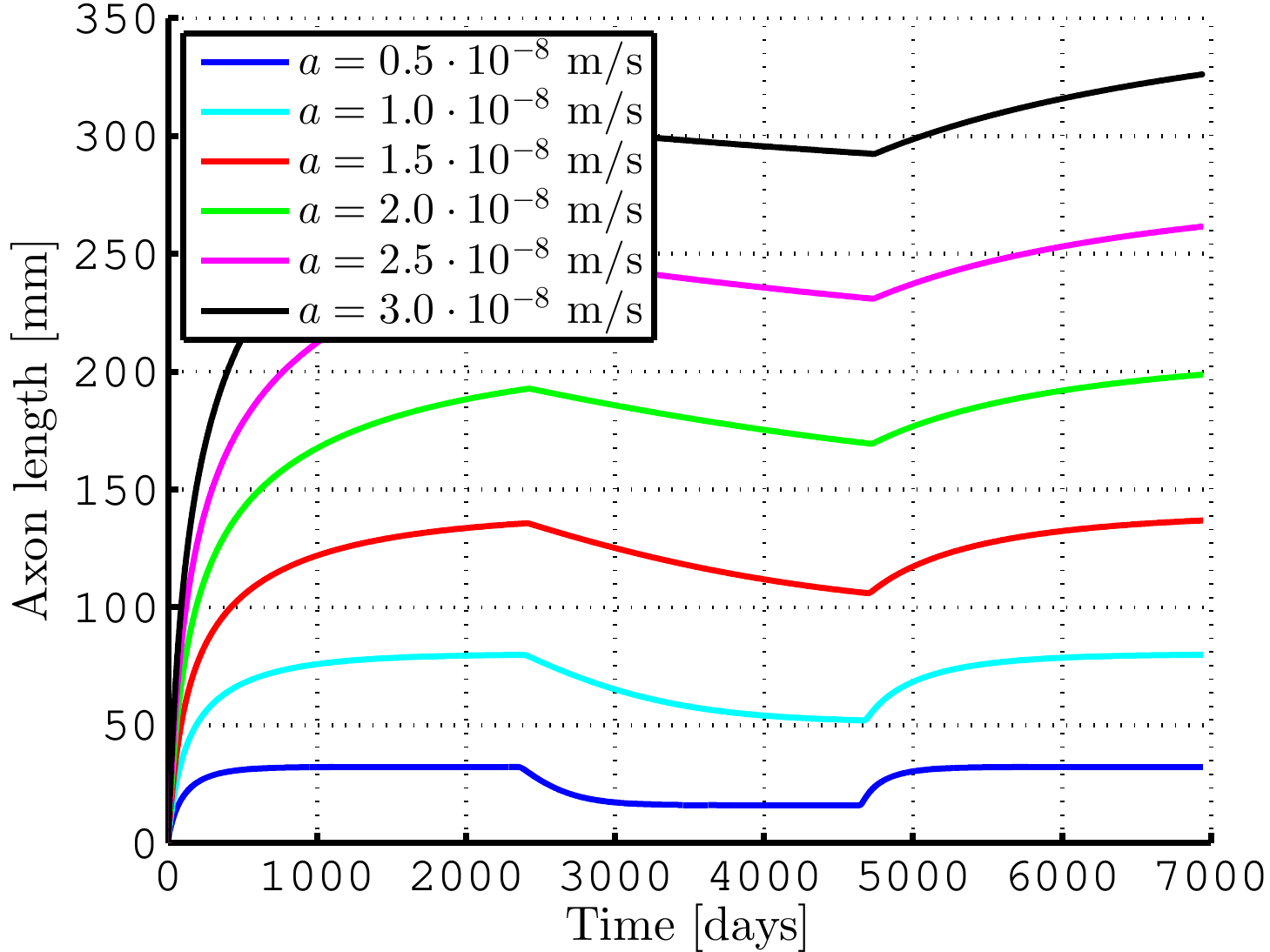}}{0mm}{0mm}
		\label{fig:a}}
\subfigure{
		\topinset{\bfseries(b)}{\includegraphics[width=.95\columnwidth]{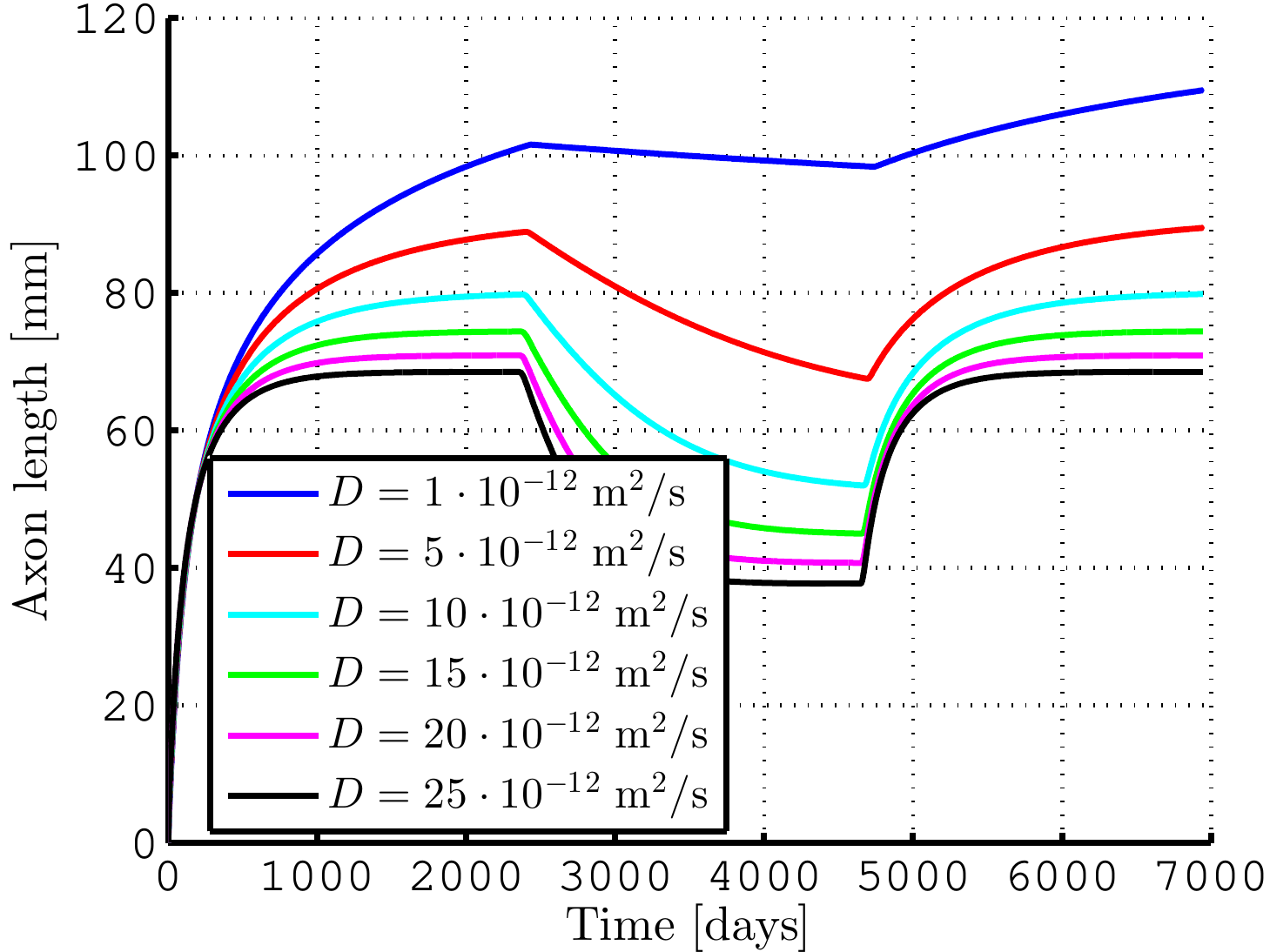}}{0mm}{0mm}
		\label{fig:D}}
\caption{Axon length $l(t)$ for varying $a$ and $D$.}
\label{fig:a_D}
\end{figure}
\begin{figure}[tb]
\centering\includegraphics[width=.95\columnwidth]{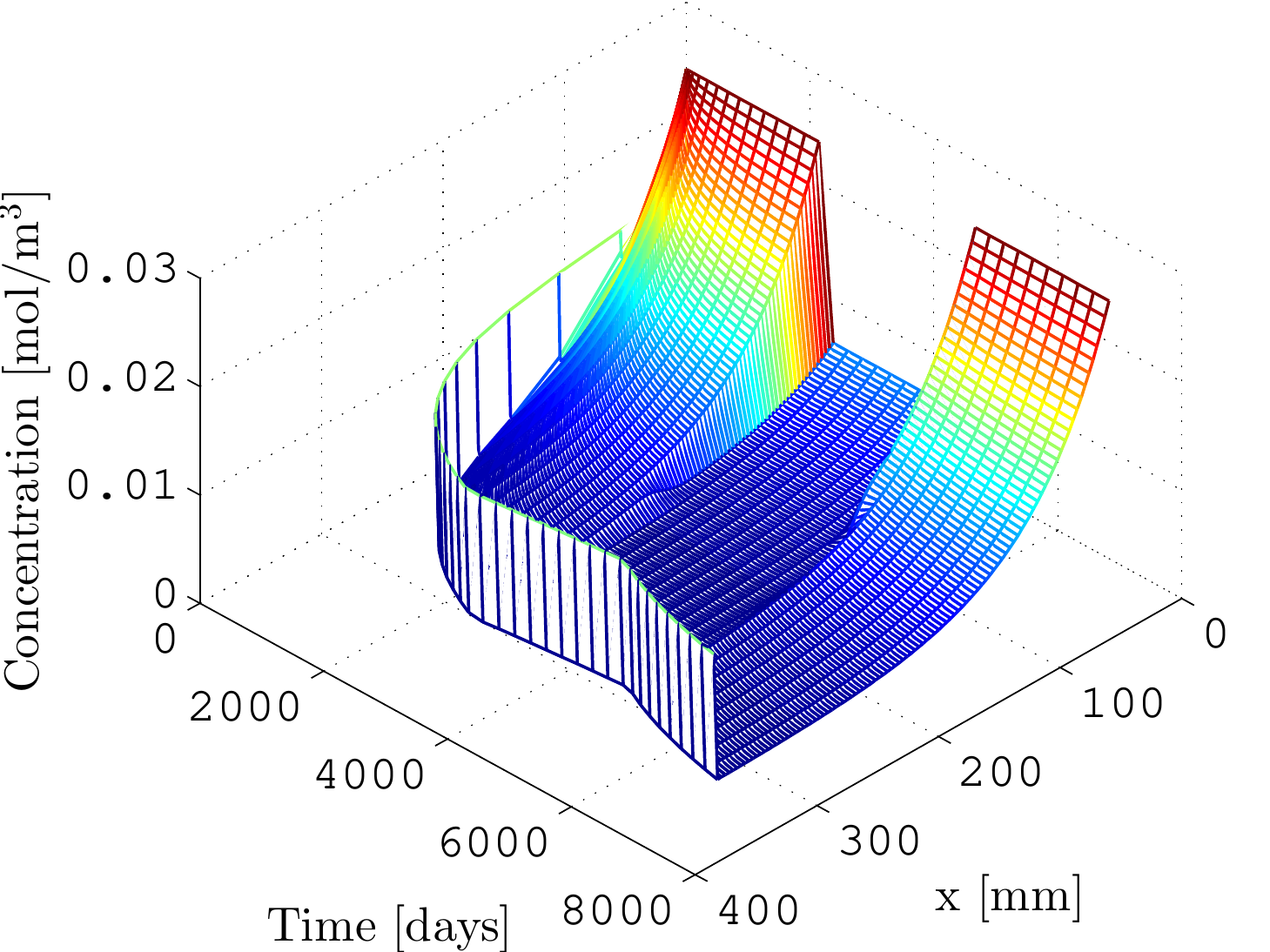}
\caption{Tubulin concentration $c(x,t)$ along the axon for $a = \SI{3.0e-8}{\metre\per\second}$. {See also Figs. \ref{fig:solution_c_slices_at_x_s_big_a} and \ref{fig:solution_c_slices_at_t_s_big_a} in Appendix B for two-dimensional slices at different values of $x$ and $t$.}}
\label{fig:a_concentration}
\end{figure}
Given the default parameter values in Table~\ref{tab:1}, each is varied and numerical simulations are performed with the settings specified in Sec.~\ref{sec:settings}. The end time is chosen large, $T = \SI{6e8}{\second} \approx 19$~years, since the convergence to steady state is of interest. The time step $\Delta \tau = \num{5e-4}$ means both fast simulations and an initial temporal error of insignificant size, cf.\ Fig.~\ref{fig:convergence}. Also note that the system \eqref{eq:model} by and large describes a parabolic problem. For such problems the initial errors typically have a small effect on the long time behaviour of the system. For the spatial resolution, $\Delta y = \num{e-3}$ suffices for all simulations except for the extreme values of the parameters $a$ and $D$.

In Fig.~\ref{fig:a} the axon length $l$ is plotted over time for varying values
of the advection velocity $a$. We observe that the axon length is sensitive to
variations in the active transport $a$ with the length increasing with the
transport velocity. However, the profile of the growth is rather insensitive to
these variations. In Fig.~\ref{fig:a_concentration} the tubulin concentration
along the axon is plotted as a function of time and space for $a =
\SI{3.0e-8}{\metre\per\second}$. Observe the sharp gradient close to the growth
cone, compare with Fig.~\ref{fig:solution_c} where $a =
\SI{1.0e-8}{\metre\per\second}$ and with the concentration profiles in
\citet[Fig.~7(a)]{SDJTB1}.

In Fig.~\ref{fig:D} we observe that $l$ grows longer for small values of the
diffusion coefficient $D$ and that the axon length is not as sensitive to
variations in $D$ as it is to variations in $a$. However, the growth profile is
more sensitive to changes in $D$: for large values of $D$ the length reaches
values close to its steady state faster than it does for smaller $D$. This can
best be seen during and after the drop in soma concentration $\cs$. For small
values of $D$, just as for big values of $a$, a sharp gradient in $c$ is
quickly created close to the growth cone, cf.\ \citet[Fig.~7]{SDJTB1}. To
resolve these gradients an extremely fine spatial resolution is needed,
therefore we ran all the simulations presented in Fig.~\ref{fig:a_D} with
$\Delta y = \num{e-4}$ except for the two biggest values of $a$ and the two
smallest values of $D$ where even $\Delta y = \num{e-5}$ is needed. If we do
not have such fine spatial resolution the axon continues its growth far past
its steady-state length. We consider that to be an unrealistic solution.

\begin{figure}[tb]
\centering\includegraphics[width=.95\columnwidth]{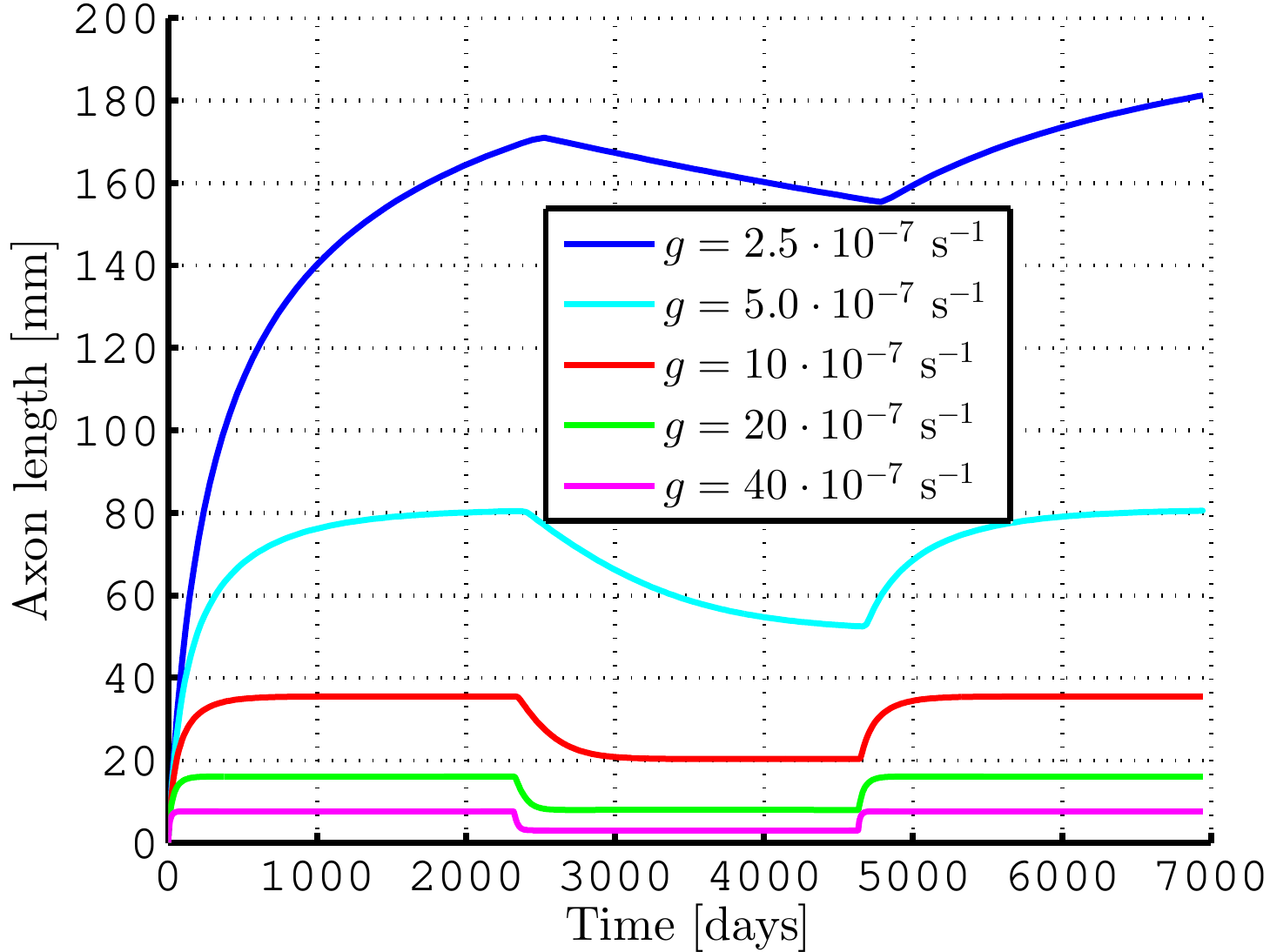}
\caption{Axon length $l(t)$ for varying degradation rate $g$.}
\label{fig:g}
\end{figure}
Fig.~\ref{fig:g} shows that the axon length is sensitive to changes in the degradation rate $g$.

\begin{figure}[tb]
\centering
\subfigure{
		\topinset{\bfseries(a)}{\includegraphics[width=.95\columnwidth]{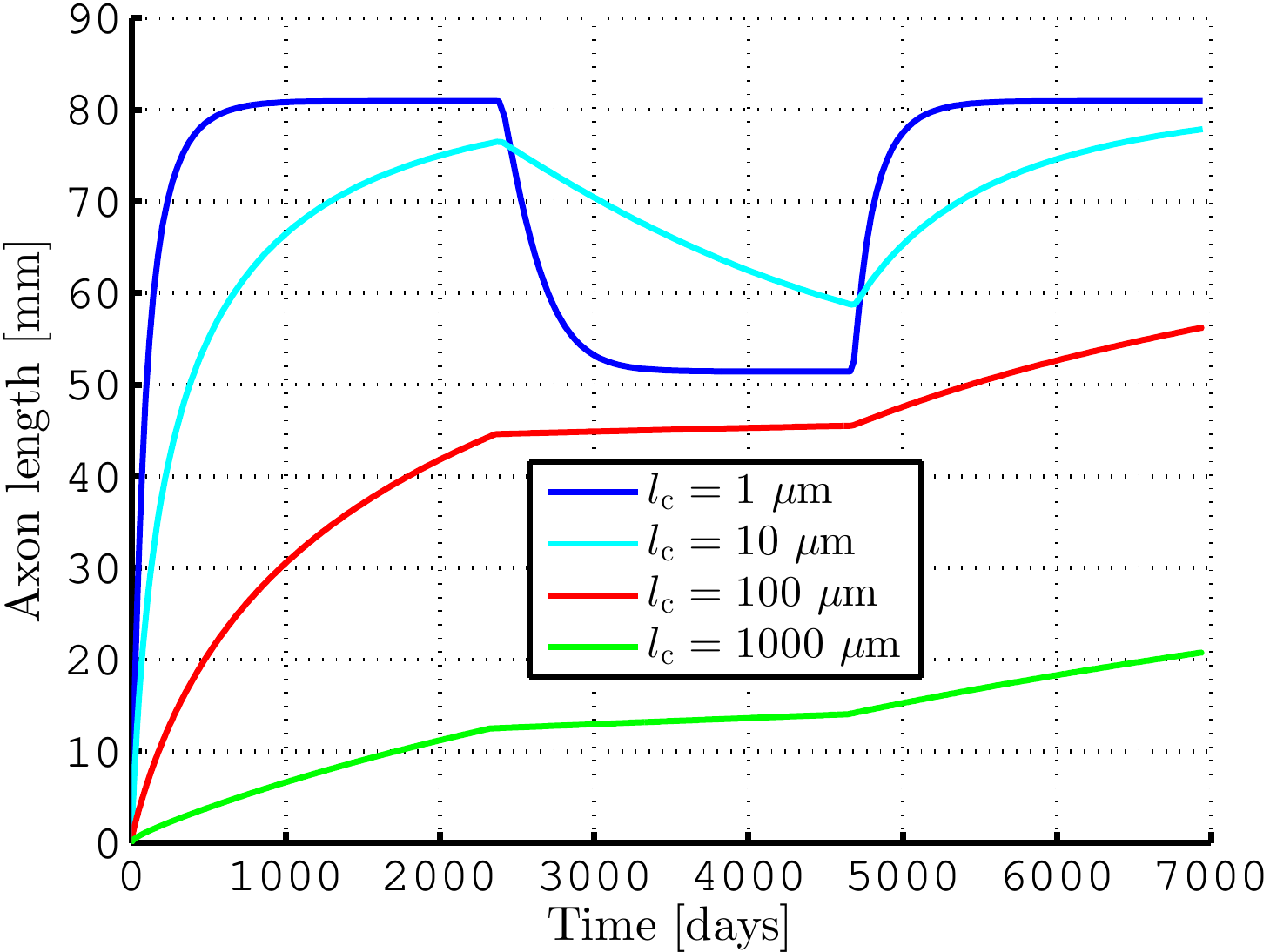}}{0mm}{0mm} 
		\label{fig:lc}}
\subfigure{
		\topinset{\bfseries(b)}{\includegraphics[width=.95\columnwidth]{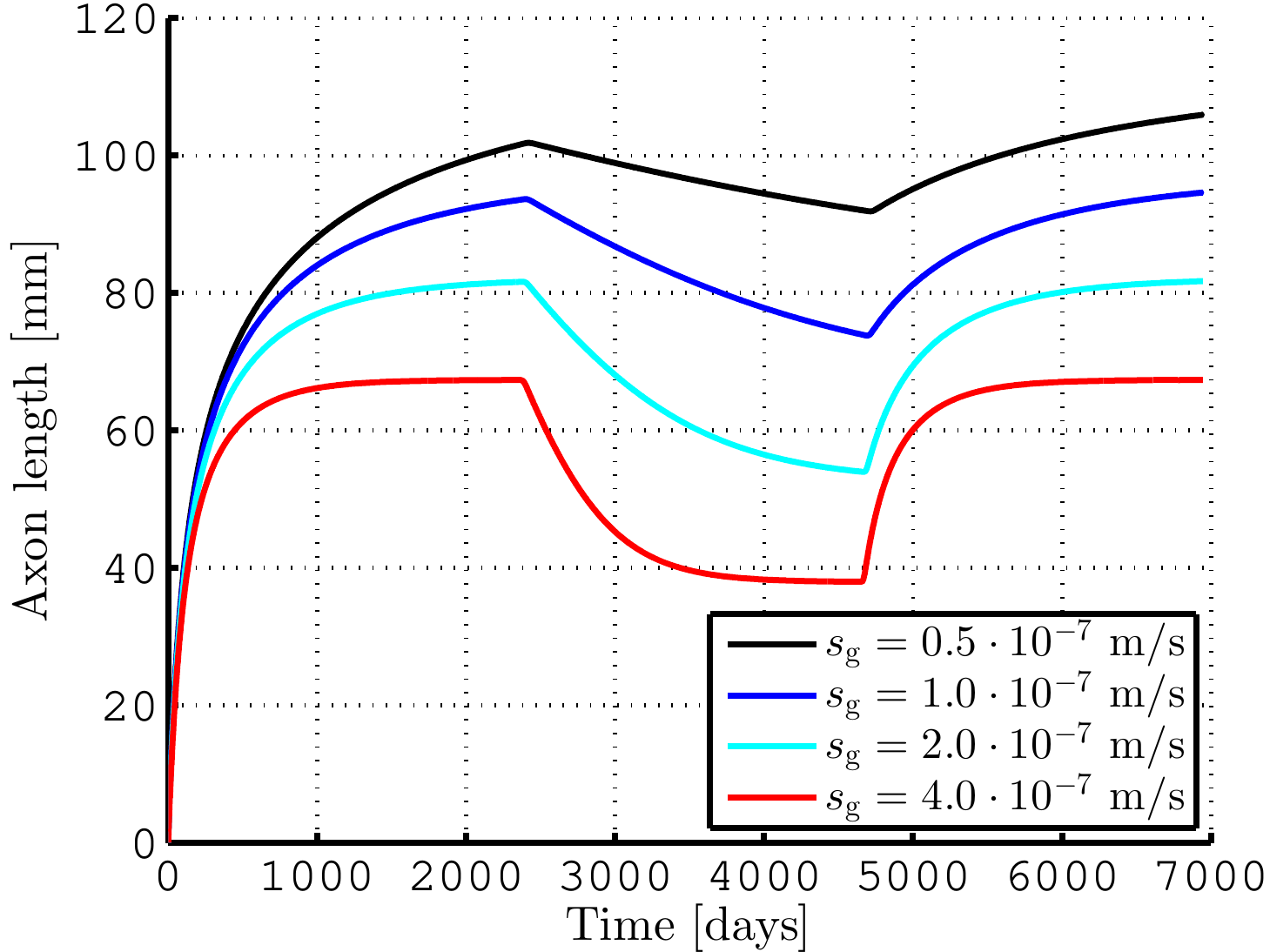}}{0mm}{0mm}
		\label{fig:sg}}
\caption{Axon length $l(t)$ for: (a) varying $\lc$ (with $\sg$ constant at its nominal value, cf.\ Table~\ref{tab:1}) and (b) $\sg$ and $\cinf$ varying together according to $\cinf = \sg/\rg$ (with $\lc$ and $\rg$ nominal).}
\label{fig:lc_sg}
\end{figure}
Recall the relation between the parameters $\lc$, $\rg$, and $\cinf$ as well as
the relation between $\rgtilde$, $\rg$, and $\cinf$ defined by the formulas
\eqref{eq:dldt}--\eqref{eq:cinf}.
In the simulations presented in Fig.~\ref{fig:lc} the parameter $\lc$ is varied
without changing $\cinf$ and $\rg$ (this means that $\sgtilde \kappa$ is varied
with $\lc$ to keep $\sg$ constant). From \citet[Fig.~8(b)]{SDJTB1} we know that
the steady-state length is insensitive to changes in the growth cone length
parameter $\lc$. This is in accordance with Fig.~\ref{fig:lc} where we also can
see that the growth speed is sensitive to these variations; the axon grows
close to its steady-state length only for the two smallest values of $\lc$.

In Fig.~\ref{fig:sg} we instead keep $\lc$ constant and let the maximum speed
of shrinkage $\sg$ vary. This affects the model \eqref{eq:model} through the
relation $\cinf = \sg/\rg$ (where $\rg$ is kept constant). We can observe that
the model indeed reacts faster to changes in soma concentration for larger
values of $\sg$. Further, for values of $\sg \gtrsim
\SI{4.3e-7}{\metre\per\second}$ we have $\cs(t)/\cinf < 1$ even for small
values of $t$. This means that we cannot expect any outgrowth for these values
of $\sg$, cf.\ \citet[Thm~4.1 and Fig.~11]{SDJTB1}.

\begin{figure}[!tb]
\centering
\subfigure{
		\topinset{\bfseries(a)}{\includegraphics[width=.95\columnwidth]{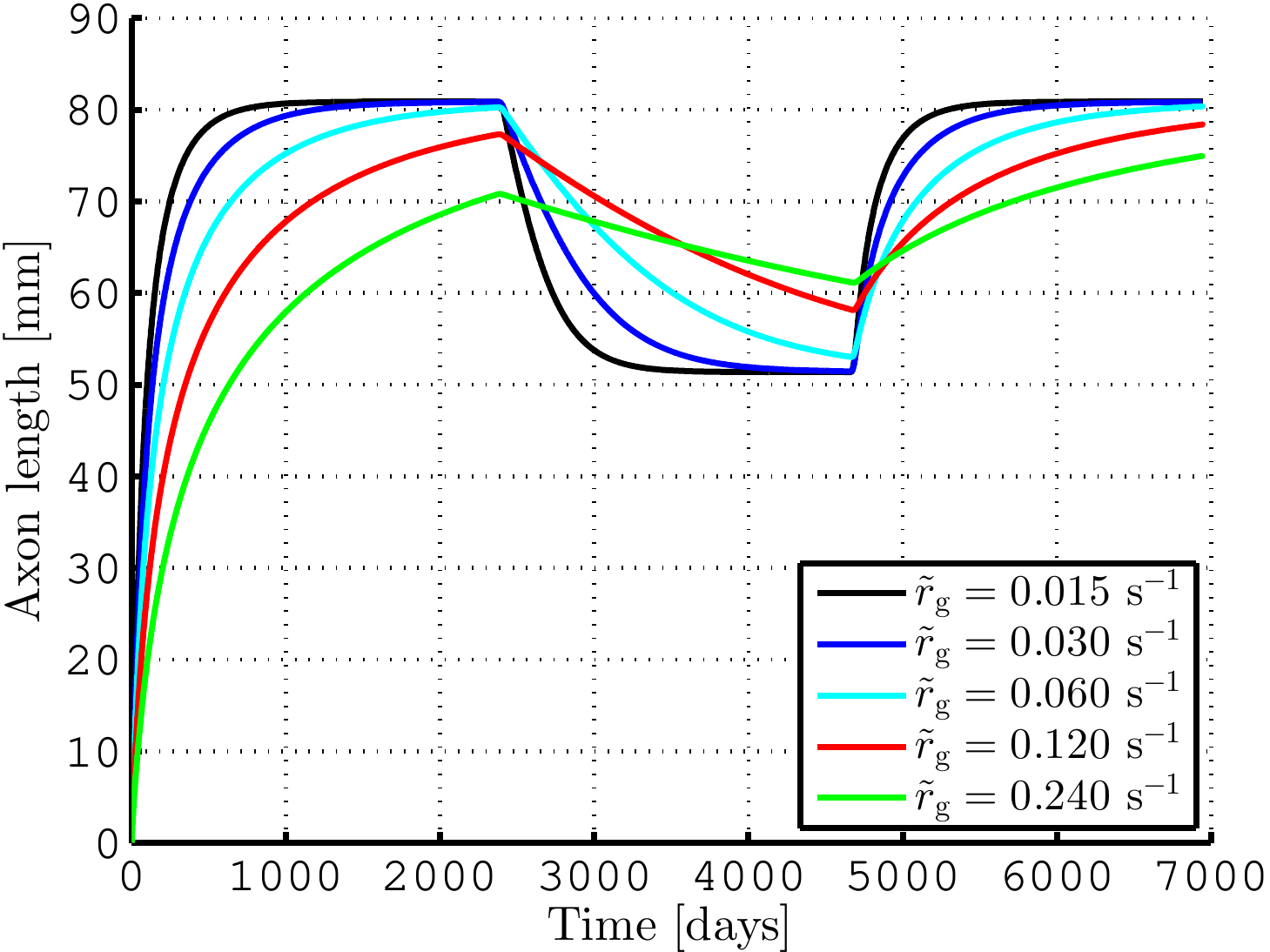}}{0mm}{0mm} 
		\label{fig:rgt}}
\subfigure{
		\topinset{\bfseries(b)}{\includegraphics[width=.95\columnwidth]{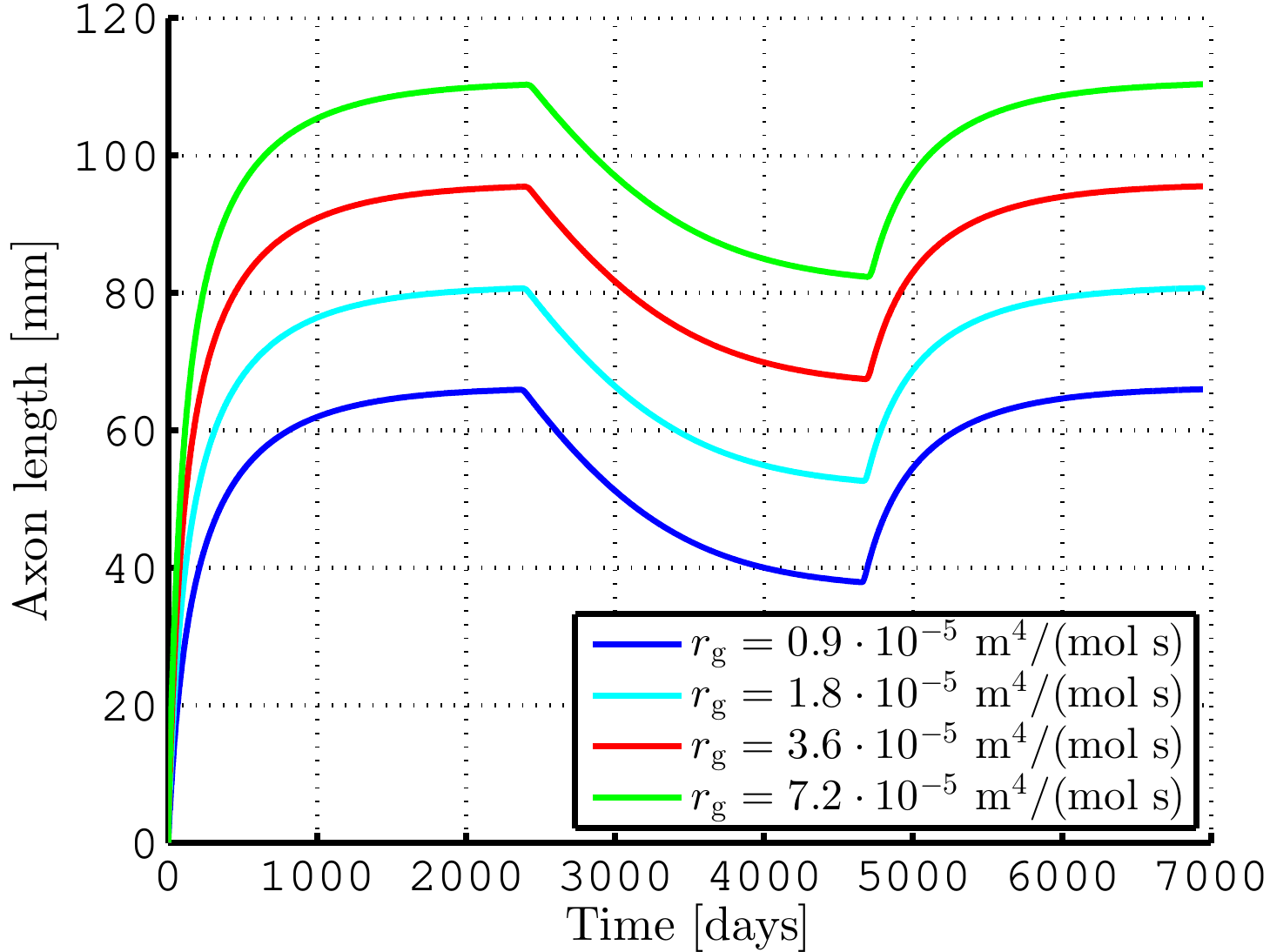}}{0mm}{0mm}
		\label{fig:rg_wo_var_rgt}}
\subfigure{
		\topinset{\bfseries(c)}{\includegraphics[width=.95\columnwidth]{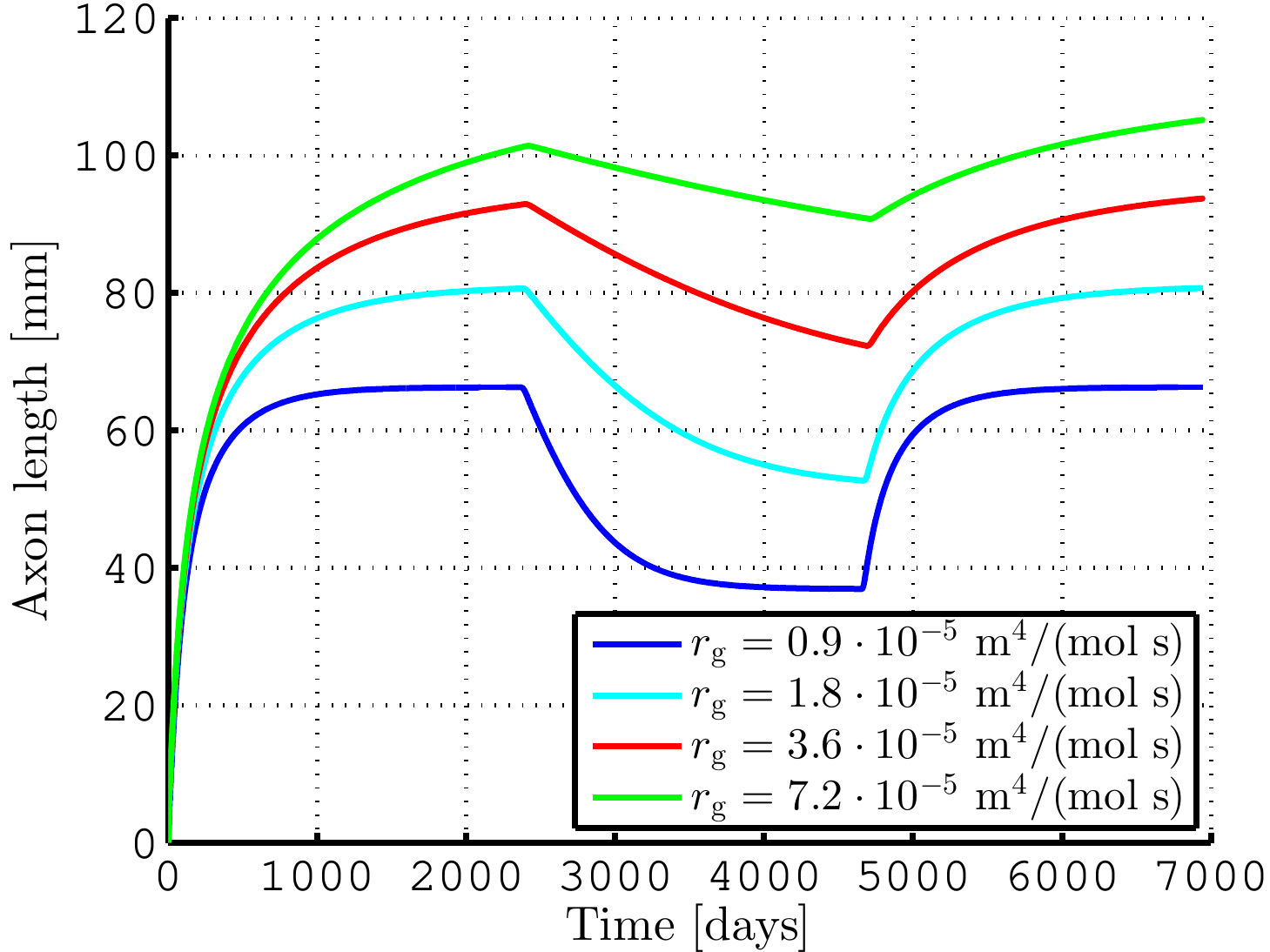}}{0mm}{0mm}
		\label{fig:rg}}
\caption{Axon length $l(t)$ for: (a) varying $\rgtilde$ (with $\rg$, $\sg$, and $\cinf$ nominal), (b) $\rg$ and $\cinf$ varying together according to \eqref{eq:cinf} (with $\rgtilde$ and $\sg$ nominal), and (c) $\rg$, $\rgtilde$, and $\cinf$ varying together according to \eqref{eq:dldt2} and \eqref{eq:cinf} (with $\sg$ nominal).}
\label{fig:rgt_rg}
\end{figure}
In Fig.~\ref{fig:rgt_rg} the effects of changing the polymerization rates
$\rgtilde$ and $\rg$ are investigated while keeping $\sg$ constant. We can
observe that the steady-state length is sensitive to $\rg$ but unaffected by
changes in $\rgtilde$ (as was already known from \citet{SDJTB1}). The opposite
dependency seems to apply for the profile of the axon growth with $\rgtilde$
affecting the shape while $\rg$ does not. Also note that for $\rg \lesssim
\SI{8.9e-6}{\metre^4\per(\mol~~\second)}$ we have $\cs(t)/\cinf < 1$ so we
cannot expect any outgrowths for these values of $\rg$.

\begin{figure}[!tb]
\centering\includegraphics[width=.95\columnwidth]{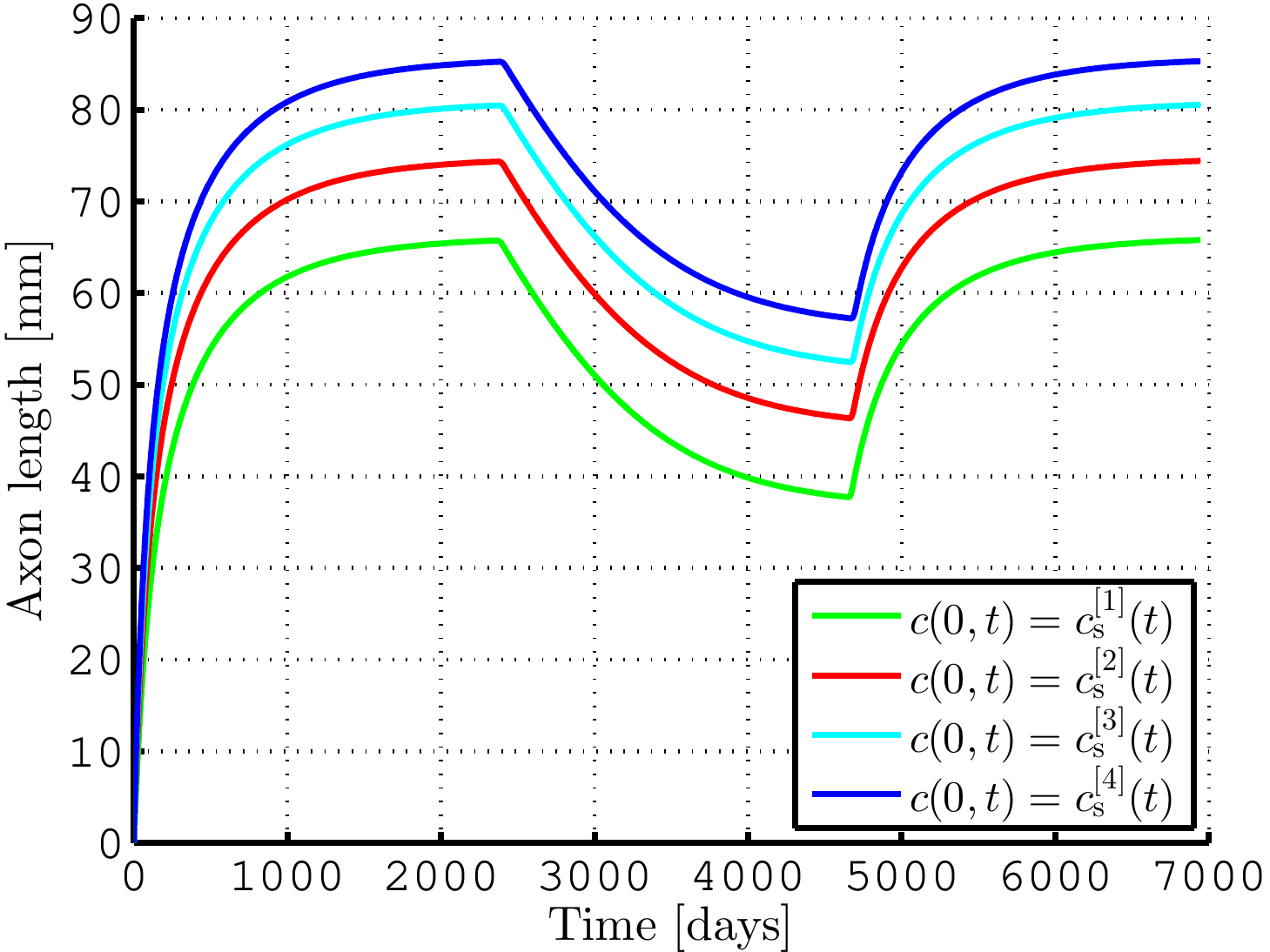}
\caption{Axon length $l(t)$ for different functions defining the left-hand side boundary condition (the soma concentration).}
\label{fig:cs0}
\end{figure}
For the simulations presented in Fig.~\ref{fig:cs0} different functions
defining the soma concentration are used. To this end, we let $\cs(t)$ be
defined by \eqref{eq:cs} and set
\[ \begin{aligned}
&\cs^{[1]}(t) := 0.5\cdot \cs(t),\quad &&\cs^{[2]}(t) := 0.75\cdot \cs(t), \\
&\cs^{[3]}(t) := \cs(t),\quad &&\cs^{[4]}(t) := 1.25\cdot \cs(t).
\end{aligned} \]
For these particular simulations, to get continuity at $(x,t) = (0,0)$, also the initial condition must be rescaled. That is, for $k = 1,2,3,4$, we use the constant initial data $c(x,0) = \cs^{[k]}(0)$.
Fig.~\ref{fig:rgt_rg} shows that larger soma concentrations give longer axon
lengths without changing the growth profile.

Finally, we keep all parameters at their nominal values and run simulations for
a range of different values of $l^0$ between \SI{e-6}{\micro\metre} and
\SI{e3}{\micro\metre}. Since this variation has no visible effect on the
solution on the considered time scale we do not include a plot. We also note
that the initial length has marginal effect on the CPU time of the simulation;
for the shortest $l^0$ the CPU time is \SI{57}{\second} whereas for the longest
initial length it is \SI{55}{\second}.

\section{Conclusions}\label{sec:conclusions}

The model of axonal growth, consisting of two ODEs coupled to one PDE defined
on an interval, which length is one unknown of the problem, has shown to be a
challenge to simulate dynamically. After a spatial transformation of the moving
boundary, it is still not straightforward to apply any numerical method.
Standard methods cannot be used to simulate the entire axonal growth, from a
very small length to the final one several magnitudes larger, in reasonable CPU
time. Furthermore, the low accuracy of first-order methods can easily simulate
that the axon grows beyond its steady state, obtained as an exact solution of
the model equations; hence, such numerical solutions are not reliable. To
obtain reliable and efficient (fast and accurate) numerical solutions, it was
necessary to use both an additional time transformation of the equations and an
application of the Peaceman-Rachford splitting scheme.

With the efficient numerical scheme presented, investigations of the biological
and physical parameters in the model were performed. These investigations
complements those for the steady-state solution of the model made in our
previous publication \citet{SDJTB1}. The following conclusions can be made for
the dynamic behaviour:
\begin{itemize}
\item {The axon grows very fast initially for a broad range of
    parameter values. As is also known from experiments, the active
    transport is the dominant driving force over diffusion for the growth velocity
		and the final length of the axon. These findings are in
    qualitative agreement with those presented by
    \citet[Figure~4]{Graham2006}.}
\item The concentration of free tubulin in the growth cone approaches
    quickly the steady-state concentration after any change in the driving
    soma concentration $\cs(t)$.
\item The concentration profile of tubulin along the axon, during growth,
    is similar to the characteristic steady-state profile, which is
    generally decreasing with the distance from the soma to a minimum value
    close to the growth cone and then increasing rapidly just before the
    cone.
\item If the size of the growth cone is increased from 1~$\mu$m to
    10~$\mu$m, the growth velocity is decreased but the final axon length
    is about the same.
\item The numerical scheme gives the possibility for a careful examination
    of the influence of the variations of the parameters related to the
    (de)polymerization on the dynamic solution. For example, the
    polymerization rate coefficient $\rgtilde$ does not influence the final
    axon length (known from \citet{SDJTB1}), but it has a great influence on
    the growth velocity of the axon; see Fig.~\ref{fig:rgt}.
\item {In Sec.\ \ref{sec:scaling_in_time} we have described time-scaling with respect to advection $a$ \eqref{eq:gamma} and with respect to $D$ \eqref{eq:scaling_diffusion}. The scaling with $D$ performs better when the axon is short whereas the one with $a$ performs better when the axon is long, see Fig.\ \ref{fig:testing_diffusion_scaling_l}. This is in agreement with the expected behaviour of advection and diffusion at different length scales. Note that, switching between the scalings at some intermediate axon length could give a scheme with improved performance. This could be of importance if our model were to be expanded to a larger one that would be more expensive to simulate. For the studies carried out in this article, however, the usage of only scaling with $a$ has shown to be sufficient.}
\end{itemize}

{
\section*{Acknowledgements}
The authors thank the reviewers for valuable suggestions, in particular, the idea to investigate the scaling with respect to diffusion.}

\appendix

\section{Proof of Lemma~\ref{lem:A_negative_definite}} \label{app:Lemma_A_negative_definite}

We begin by noting that any positive definite matrix is invertible and that any square matrix is positive definite when its symmetric part has only positive eigenvalues. That is, the linear system of equations \eqref{eq:PR_PDE_IE_linear_system} has a unique solution when the eigenvalues of
\[ \bS(\bU^{n+\frac12},\by) := \frac{\left(\bI - \frac{\Delta\tau}{2}\bA(\bU^{n+\frac12},\by)\right) + \left(\bI - \frac{\Delta\tau}{2} \bA(\bU^{n+\frac12},\by)\right)\trans}{2} \]
are all positive. The entries of the main diagonal, respectively the super and sub diagonals are given as follows
\[ \begin{aligned}
&\begin{split}
&s_{j,j}(\bU^{n+\frac12},\by) \\
&\quad = 1 + \frac{\Delta\tau}{2} \left(\frac{2D}{a(\Delta y)^2}\frac{1}{L^{n+\frac12}} + \frac{g}{a}L^{n+\frac12}\right),
\end{split} && j = 1, \dots, M-1, \\
&\begin{split}
&s_{j,j+1}(\bU^{n+\frac12},\by) = s_{j+1,j}(\bU^{n+\frac12},\by)\\
&\quad = - \frac{\Delta\tau}{2} \left(\frac{\rg}{4a}(\Cc^{n+\frac12}-\cinf) + \frac{D}{a(\Delta y)^2}\frac{1}{L^{n+\frac12}}\right),
\end{split} && j = 1, \dots, M-2.
\end{aligned} \]
Thus, $\bS(\bU^{n+1/2},\by)$ is a symmetric, tridiagonal Toeplitz matrix meaning that the eigenvalues are given by the following formula
\[ \lambda_k = s_{j,j}(\bU^{n+\frac12},\by) + 2s_{j,j+1}(\bU^{n+\frac12},\by) \cos(k\pi\Delta y), \quad k = 1, \dots, M-1. \]
Then, any $\lambda_k$ fulfils the following inequality
\[ \begin{aligned}
\lambda_k &\geq s_{j,j}(\bU^{n+\frac12},\by) - |2s_{j,j+1}(\bU^{n+\frac12},\by) \cos(k\pi\Delta y)| \\
	&\geq s_{j,j}(\bU^{n+\frac12},\by) - |2s_{j,j+1}(\bU^{n+\frac12},\by)|.
\end{aligned} \]
By plugging in the entries and using the triangle inequality we get
\[ \begin{aligned}
\lambda_k &\geq 1 + \frac{\Delta\tau}{2} \left(\frac{2D}{a(\Delta y)^2}\frac{1}{L^{n+\frac12}} + \frac{g}{a}L^{n+\frac12}\right) \\
	&- \Delta\tau \left\lvert\frac{\rg}{4a}(\Cc^{n+\frac12}-\cinf) + \frac{D}{a(\Delta y)^2}\frac{1}{L^{n+\frac12}}\right\rvert \\
	&\geq 1 + \frac{\Delta\tau}{2} \left(\frac{2D}{a(\Delta y)^2}\frac{1}{L^{n+\frac12}} + \frac{g}{a}L^{n+\frac12}\right) \\
	&- \Delta\tau \left( \left\lvert\frac{\rg}{4a}(\Cc^{n+\frac12}-\cinf)\right\rvert + \left\lvert\frac{D}{a(\Delta y)^2}\frac{1}{L^{n+\frac12}}\right\rvert\right) \\
	&= 1 + \Delta\tau \left( \frac{g}{2a}L^{n+\frac12} - \frac{\rg}{4a}|\Cc^{n+\frac12}-\cinf| \right) \\
	&\geq 1 - \Delta\tau \frac{\rg}{4a}|\Cc^{n+\frac12}-\cinf| \ \ \geq \ \ 1 - \Delta\tau \frac{\rg}{4a}\gamma,
\end{aligned} \]
where we have used the definition of $\gamma$ given by \eqref{eq:lemma_1_solution_condition}. We conclude that the eigenvalues of $\bS(\bU^{n+1/2},\by)$ are positive when
\[ \Delta\tau < \frac{4a}{\rg} \cdot \frac{1}{\gamma}, \]
and therefore, for these values of $\Delta \tau$, the system \eqref{eq:PR_PDE_IE_linear_system} has a unique solution.

\begin{figure}[!htb]
\centering
\subfigure{
		\topinset{}{\includegraphics[width=.95\columnwidth]{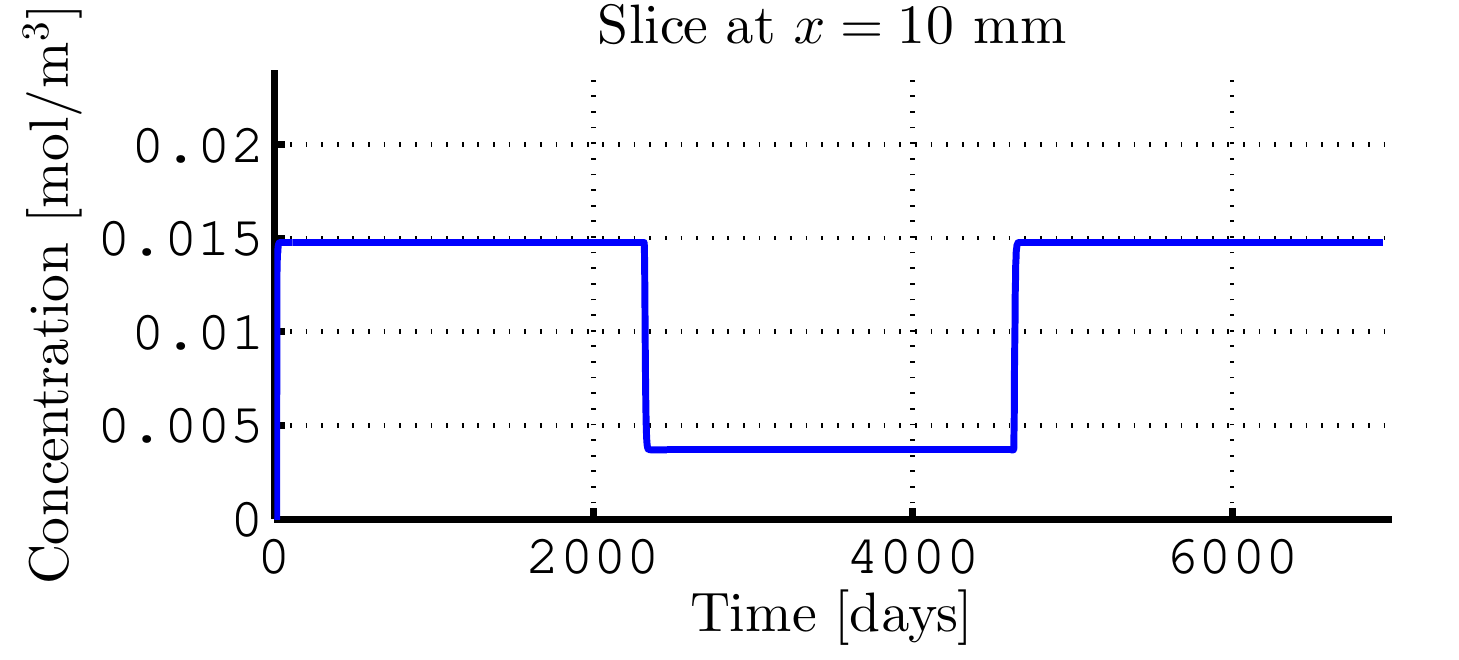}}{0mm}{6mm}
		\label{fig:solution_c_at_x=ld10}}
\subfigure{
		\topinset{}{\includegraphics[width=.95\columnwidth]{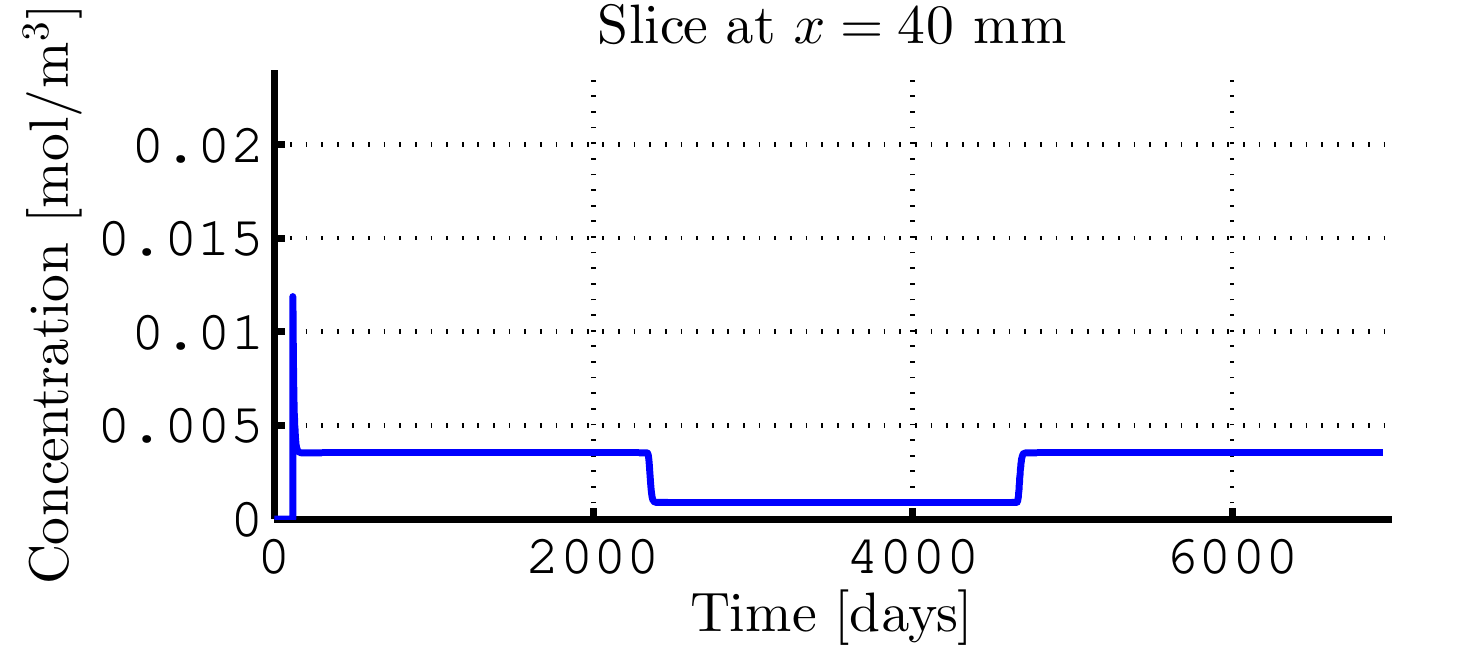}}{0mm}{0mm}
		\label{fig:solution_c_at_x=ld2}}
\subfigure{
		\topinset{}{\includegraphics[width=.95\columnwidth]{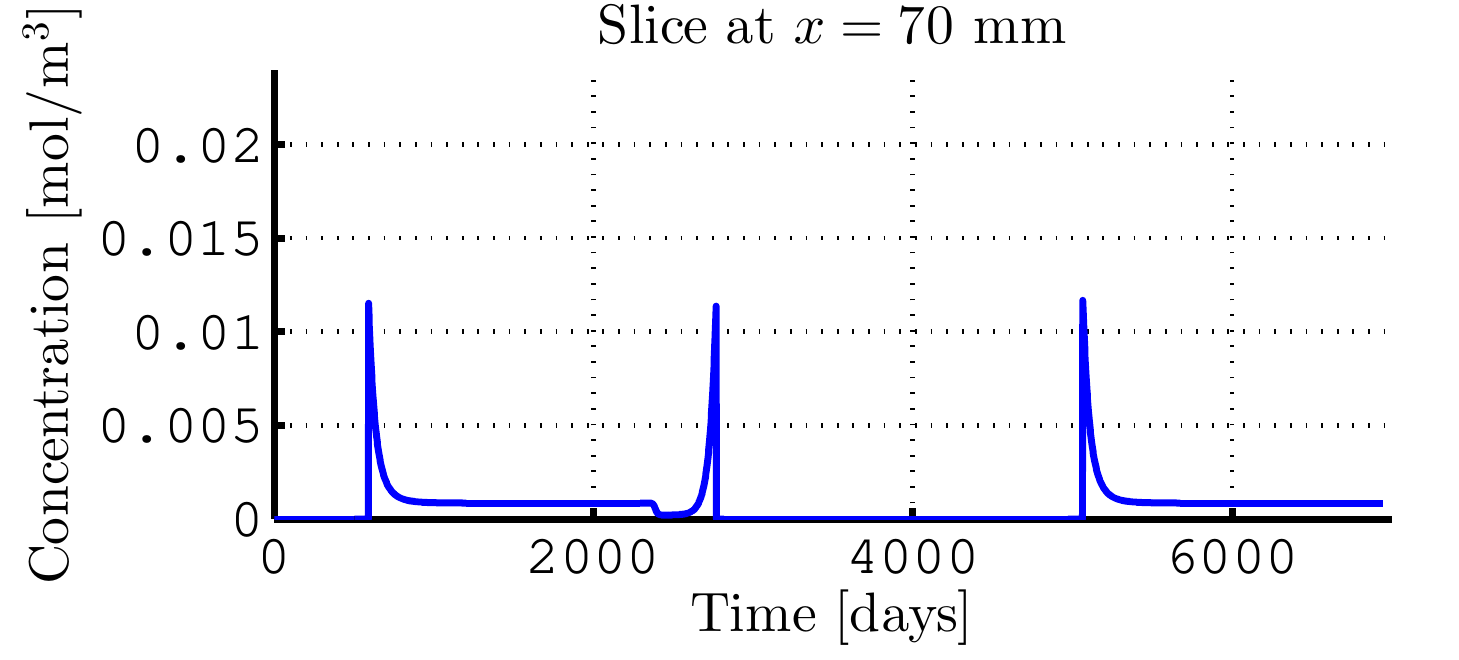}}{0mm}{0mm}
		\label{fig:solution_c_at_x=9ld10}}
\subfigure{
		\topinset{}{\includegraphics[width=.95\columnwidth]{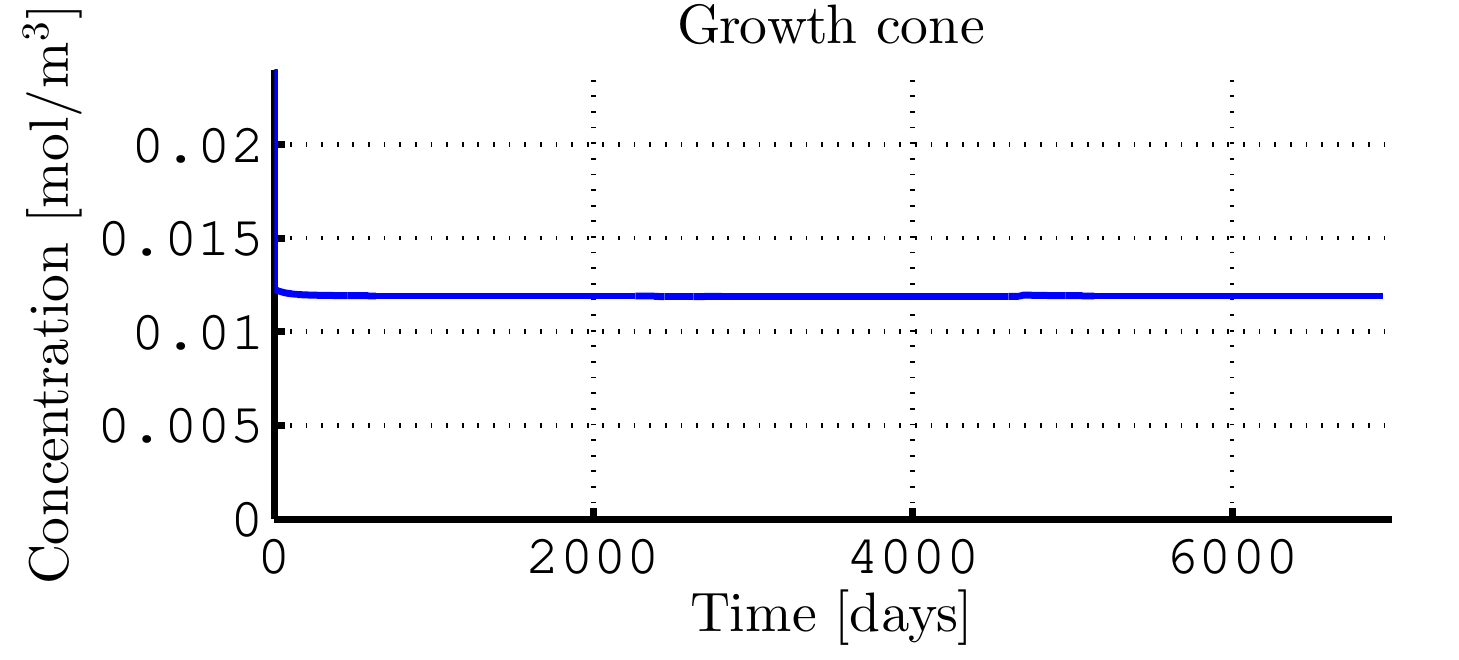}}{0mm}{0mm}
		\label{fig:solution_c_at_x=l}}
\caption{{Slices of Fig. \ref{fig:solution_c} at three different $x$ values and at the growth cone. Note that the spatial position of the latter changes with time. The spikes in the second and third subfigures correspond to times where the respective slice is close to the cone. Further, note that the concentration in the growth cone is largely unaffected by the soma concentration, instead a higher value in the latter results in a longer axon.}}
\label{fig:solution_c_slices_at_x_s}
\end{figure}

\begin{figure}[!htb]
\centering
\subfigure{
		\topinset{}{\includegraphics[width=.95\columnwidth]{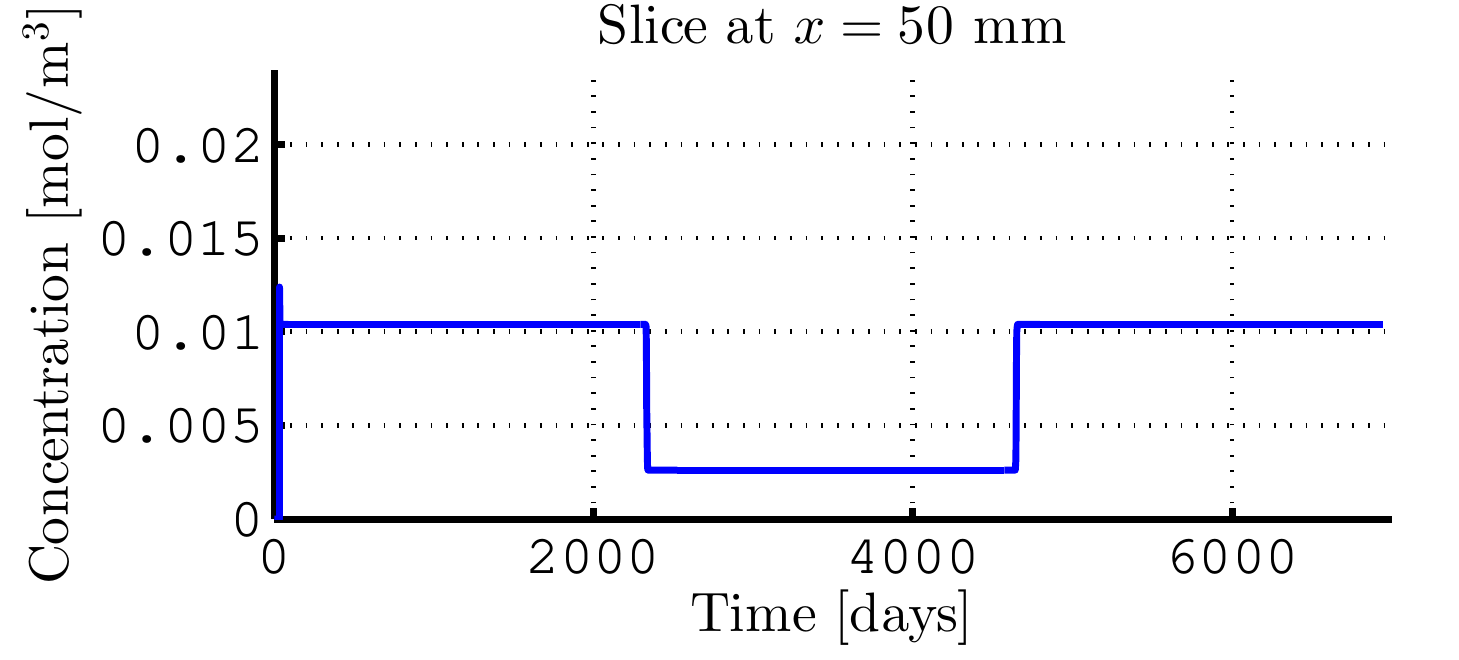}}{0mm}{6mm}
		\label{fig:solution_c_at_x=ld10_big_a}}
\subfigure{
		\topinset{}{\includegraphics[width=.95\columnwidth]{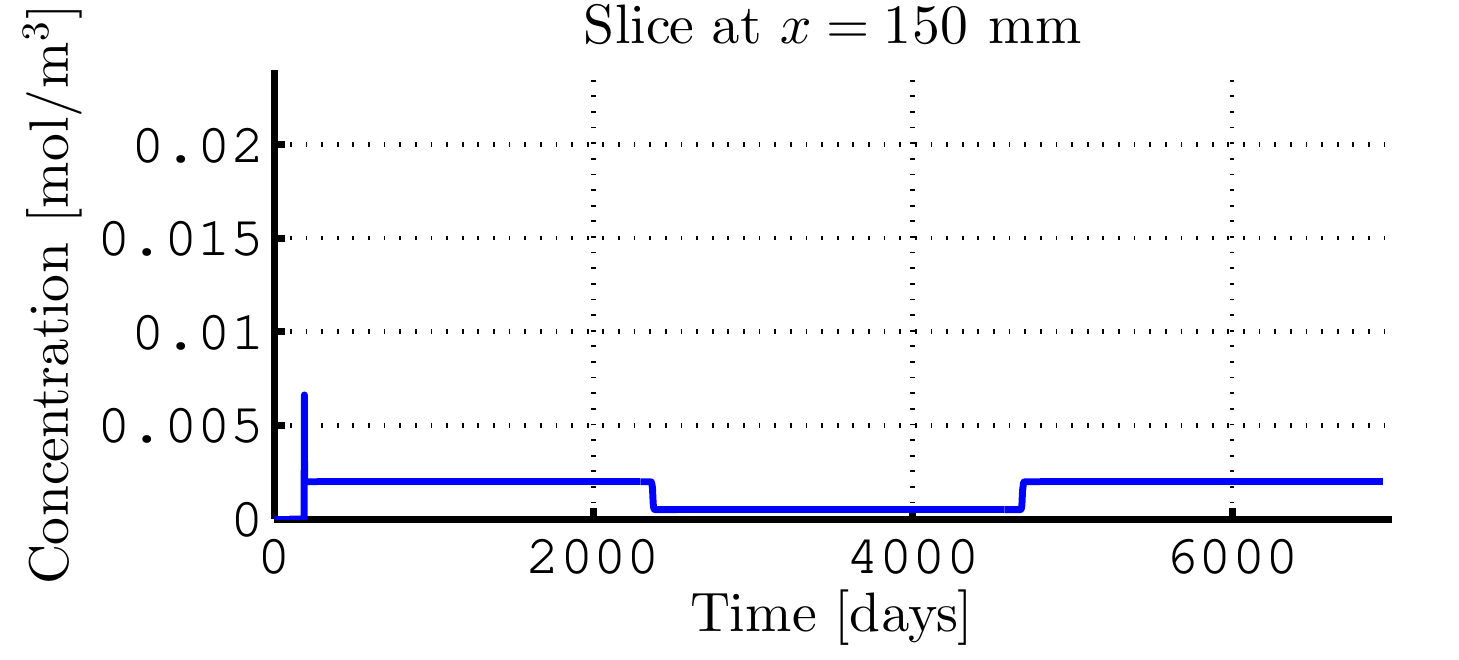}}{0mm}{0mm}
		\label{fig:solution_c_at_x=ld2_big_a}}
\subfigure{
		\topinset{}{\includegraphics[width=.95\columnwidth]{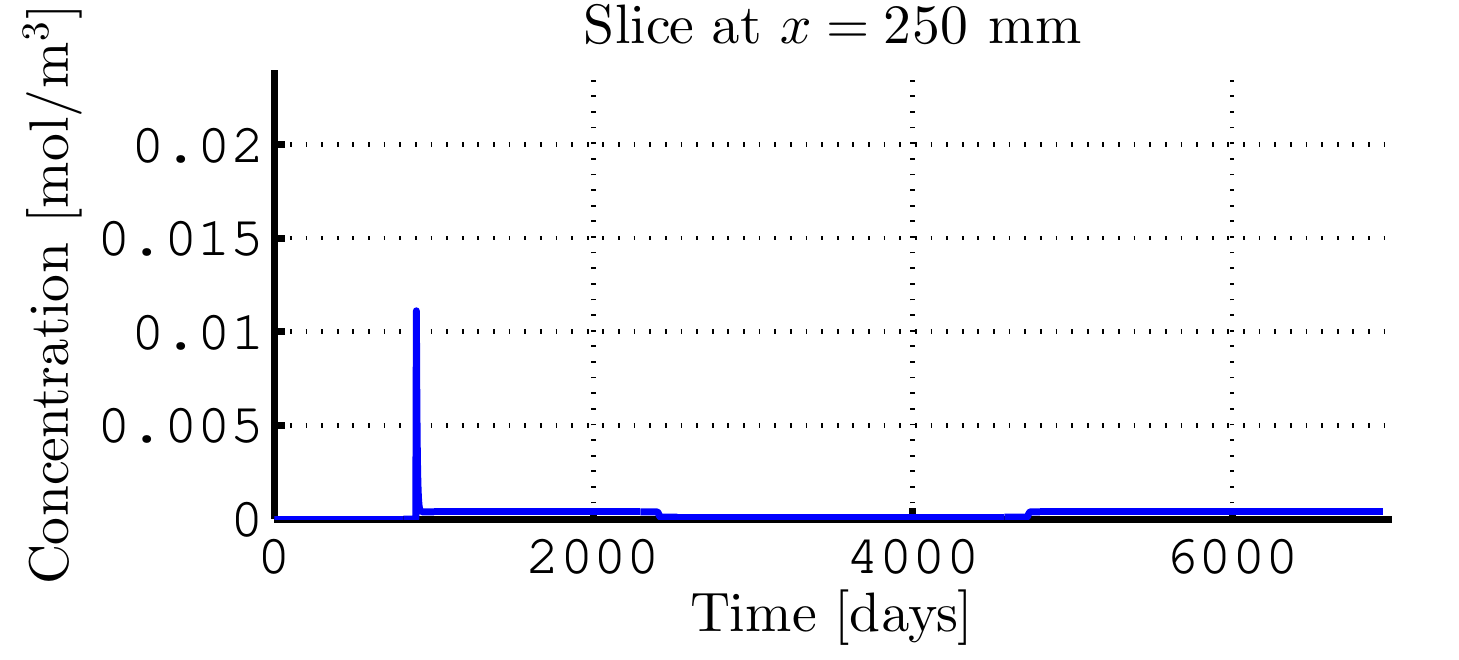}}{0mm}{0mm}
		\label{fig:solution_c_at_x=9ld10_big_a}}
\subfigure{
		\topinset{}{\includegraphics[width=.95\columnwidth]{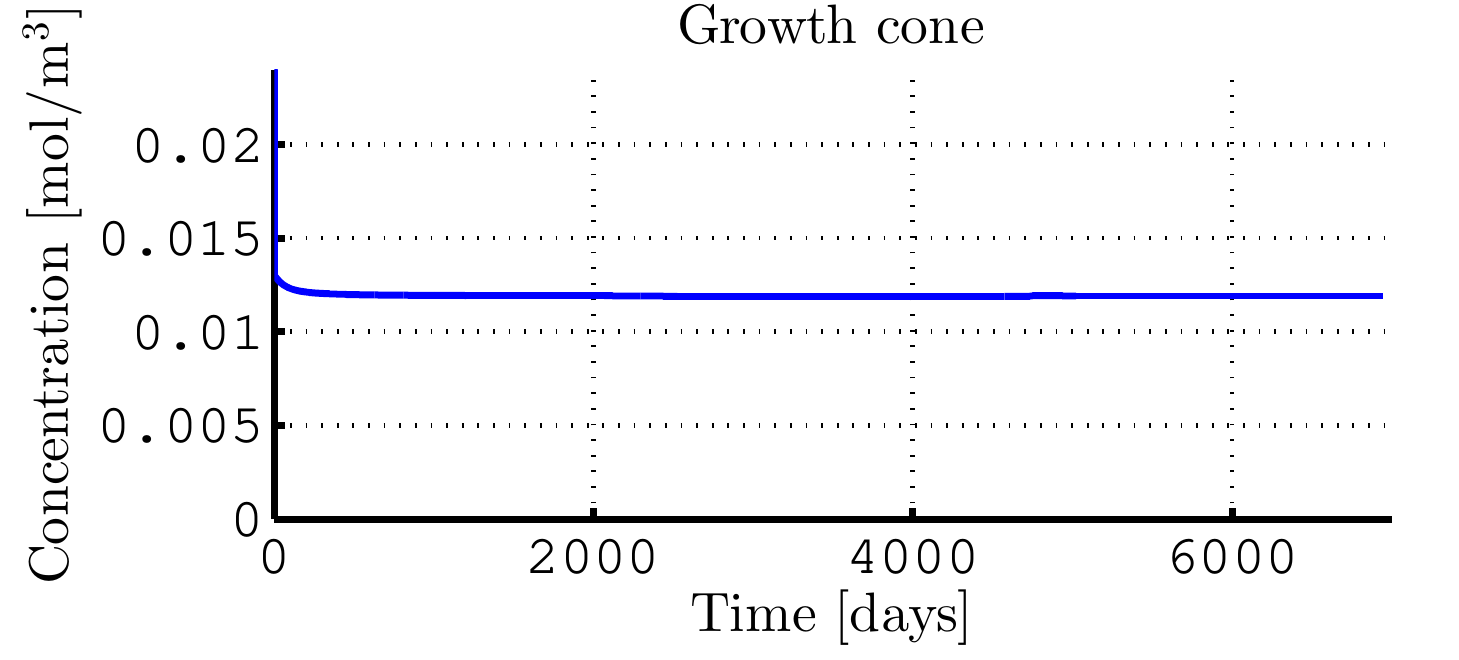}}{0mm}{0mm}
		\label{fig:solution_c_at_x=l_big_a}}
\caption{{Slices of Fig. \ref{fig:a_concentration} at three different $x$ values and at the growth cone. As the three times larger advection velocity $a$ gives a far longer axon, different $x$ values are chosen compared to those in Fig. \ref{fig:solution_c_slices_at_x_s}. Note how the higher advection velocity gives smaller concentration values in the inner of the axons; the tubulin is concentrated close to the soma and the growth cone.}}
\label{fig:solution_c_slices_at_x_s_big_a}
\end{figure}

\begin{figure}[!htb]
\centering
\subfigure{
		\topinset{}{\includegraphics[width=.95\columnwidth]{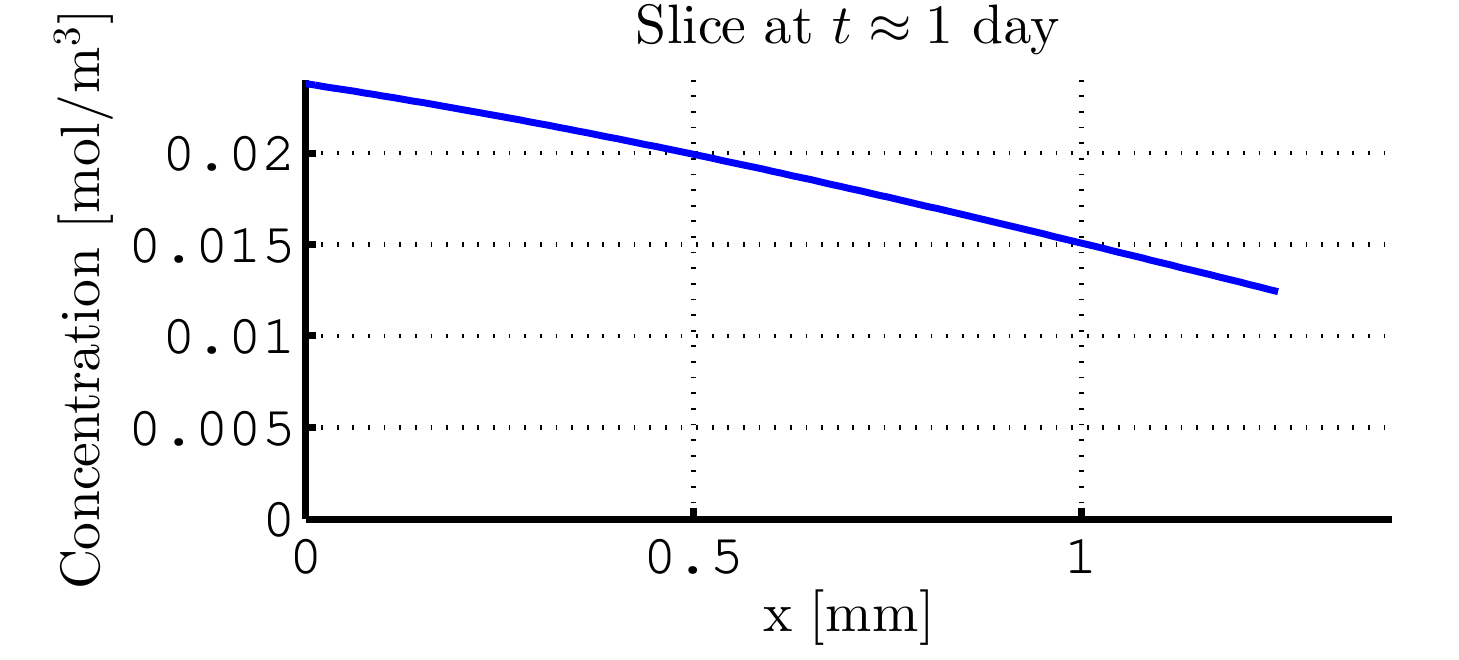}}{0mm}{6mm}
		\label{fig:solution_c_at_t=1day}}
\subfigure{
		\topinset{}{\includegraphics[width=.95\columnwidth]{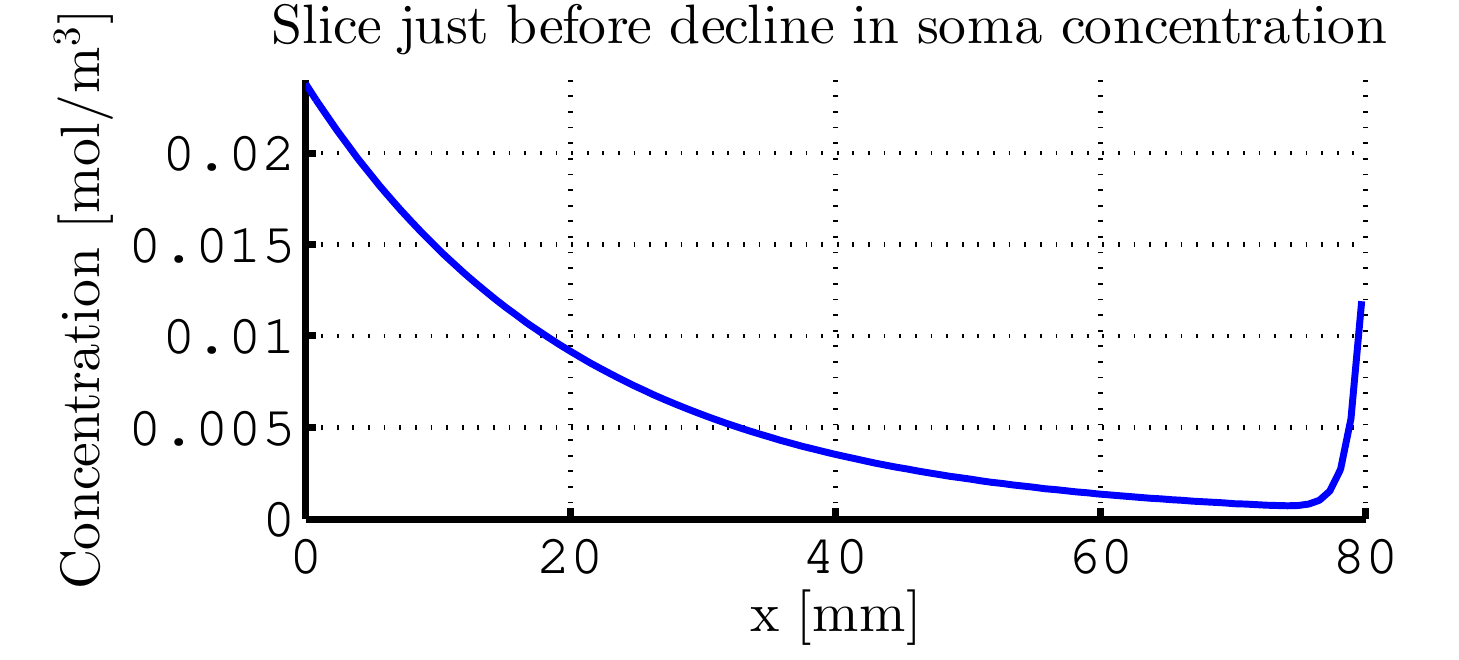}}{0mm}{0mm}
		\label{fig:solution_c_at_t=drop}}
\subfigure{
		\topinset{}{\includegraphics[width=.95\columnwidth]{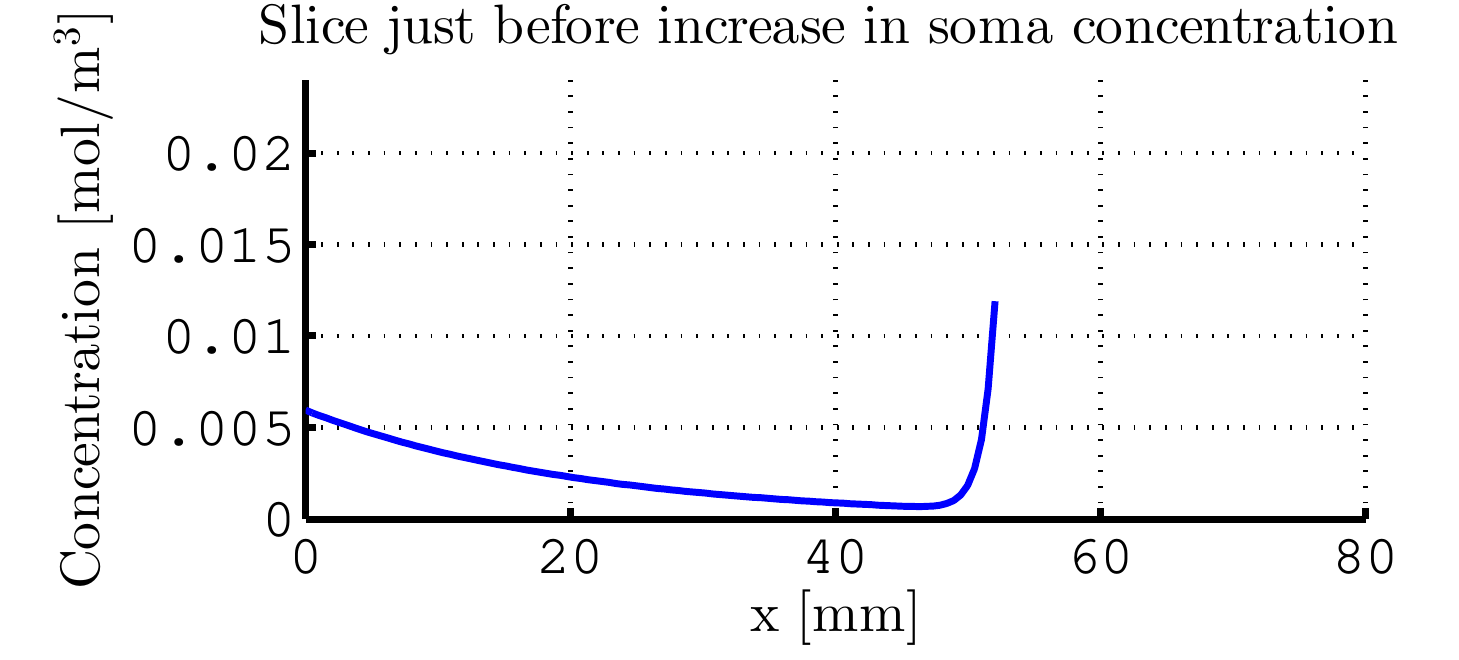}}{0mm}{0mm}
		\label{fig:solution_c_at_t=increase}}
\subfigure{
		\topinset{}{\includegraphics[width=.95\columnwidth]{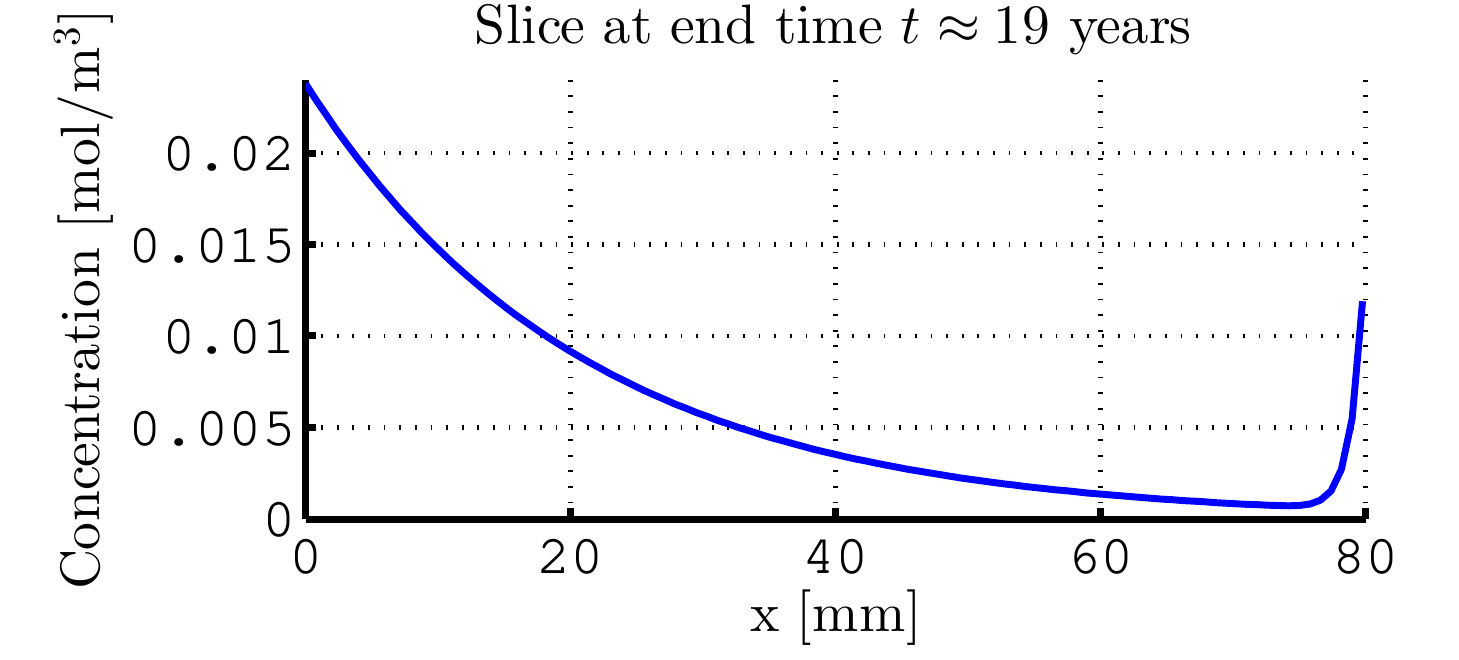}}{0mm}{0mm}
		\label{fig:solution_c_at_t=end}}
\caption{{Slices of Fig. \ref{fig:solution_c} at four different $t$ values. See also \eqref{eq:cs} for the times of decline and increase in soma concentration. Note the characteristic profile of the concentration along the axon, cf.\ \citet[Sec.~4]{SDJTB1}.}}
\label{fig:solution_c_slices_at_t_s}
\end{figure}

\begin{figure}[!htb]
\centering
\subfigure{
		\topinset{}{\includegraphics[width=.95\columnwidth]{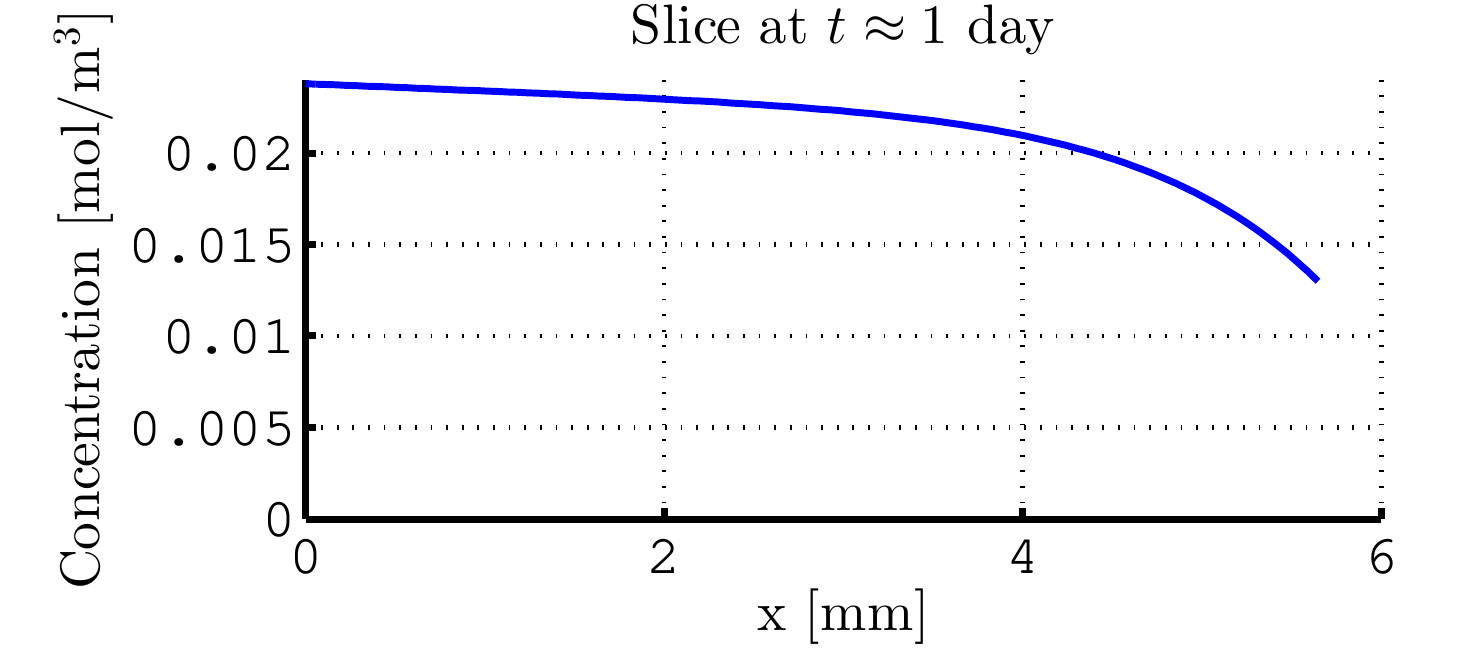}}{0mm}{6mm}
		\label{fig:solution_c_at_t=1day_big_a}}
\subfigure{
		\topinset{}{\includegraphics[width=.95\columnwidth]{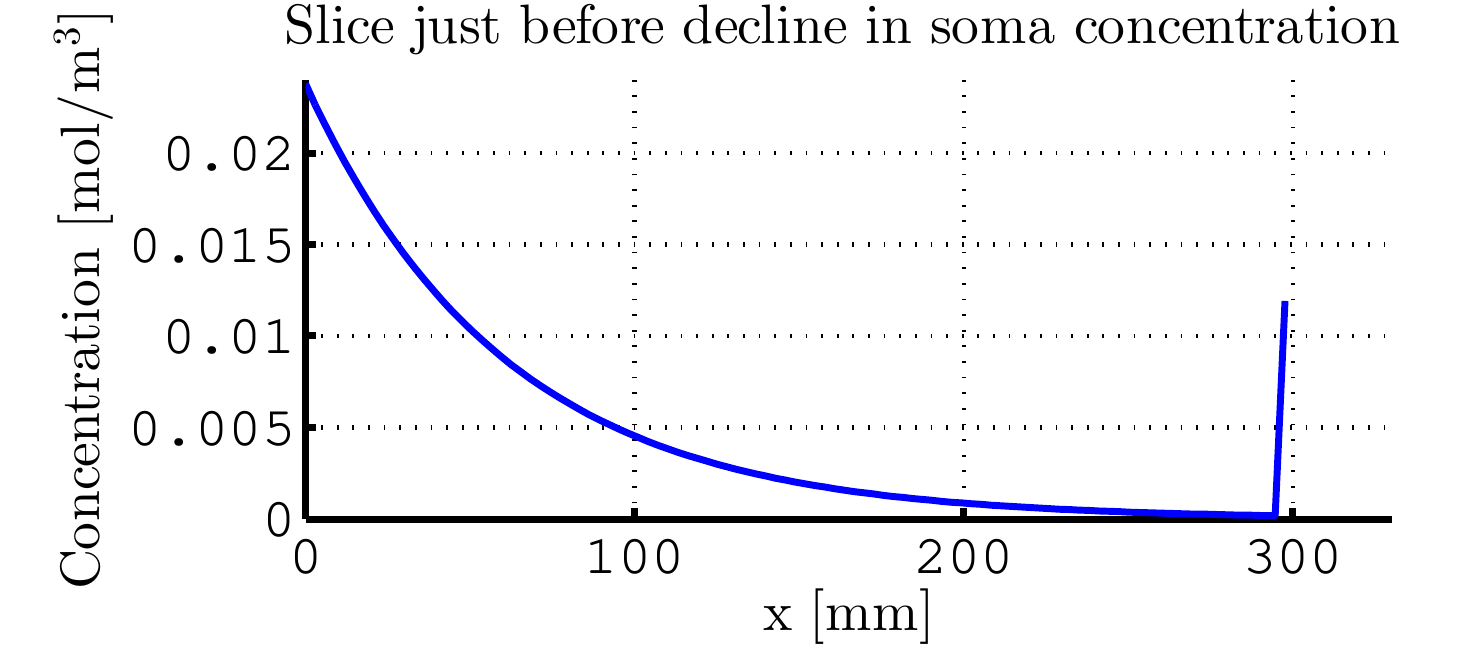}}{0mm}{0mm}
		\label{fig:solution_c_at_t=drop_big_a}}
\subfigure{
		\topinset{}{\includegraphics[width=.95\columnwidth]{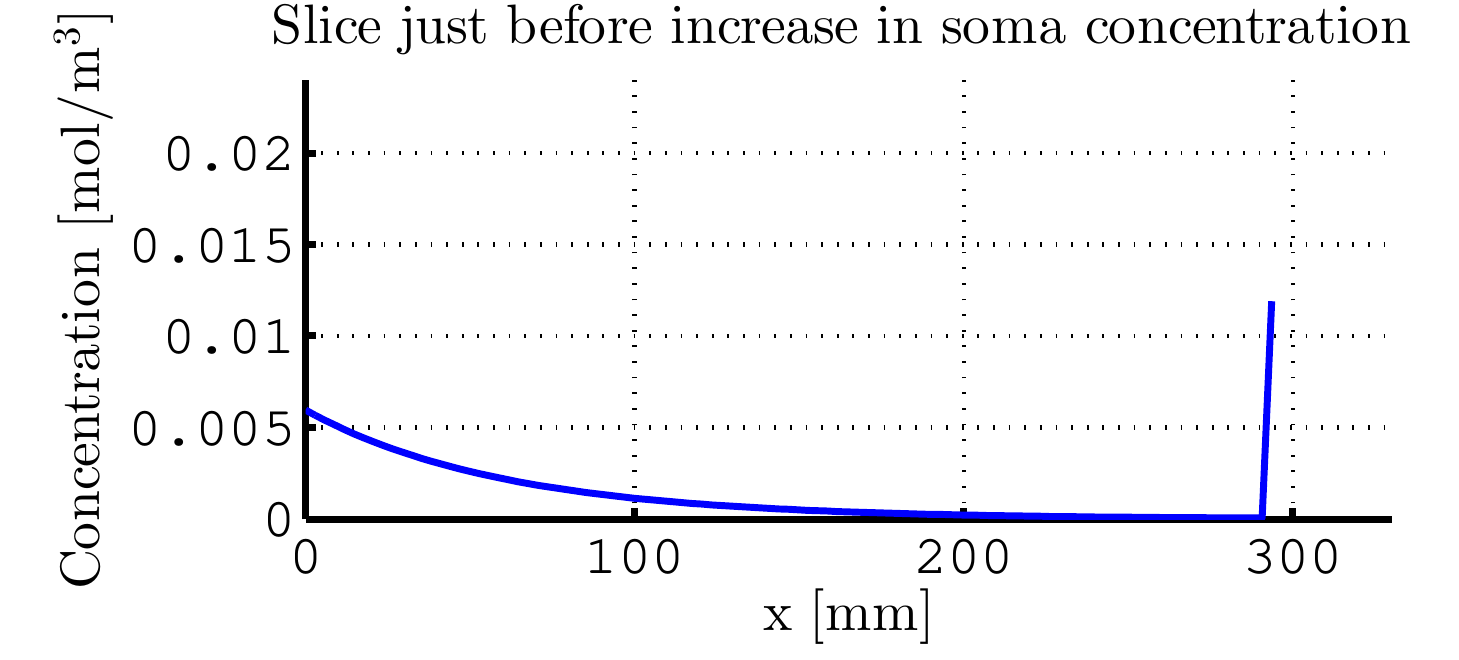}}{0mm}{0mm}
		\label{fig:solution_c_at_t=increase_big_a}}
\subfigure{
		\topinset{}{\includegraphics[width=.95\columnwidth]{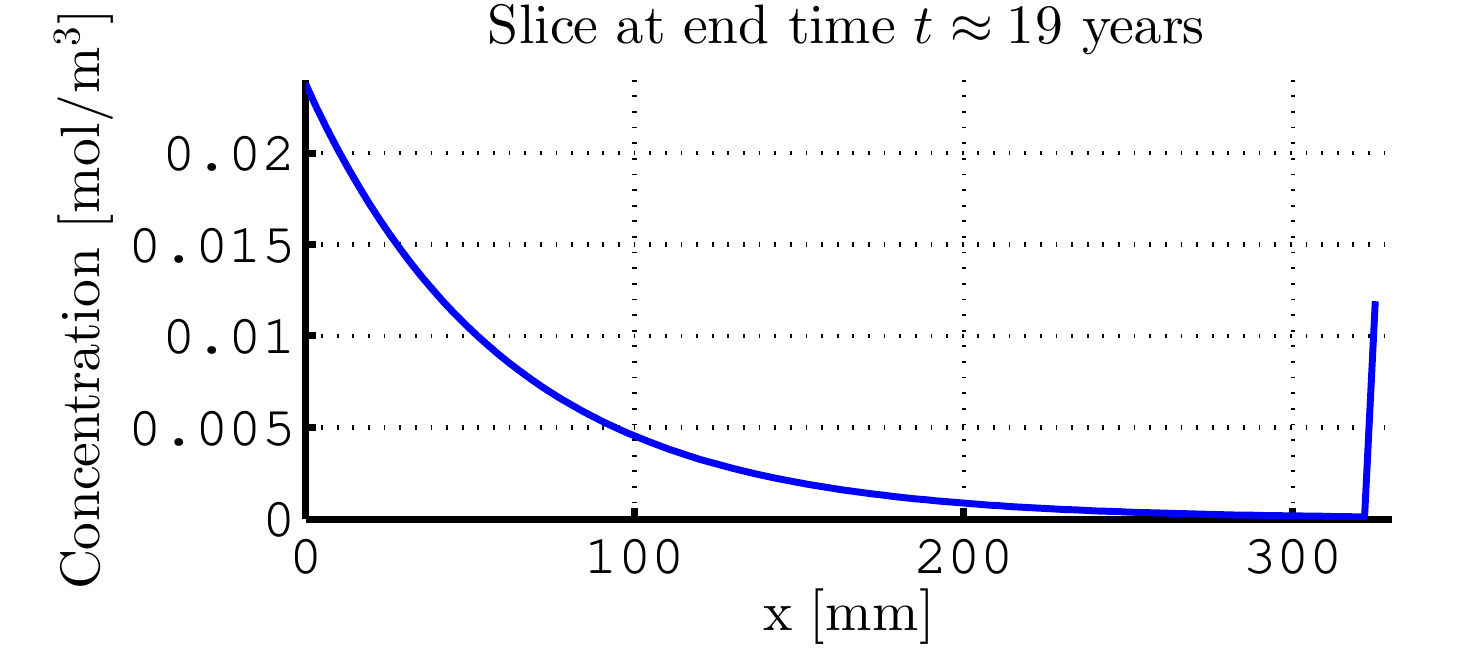}}{0mm}{0mm}
		\label{fig:solution_c_at_t=end_big_a}}
\caption{{Slices of Fig. \ref{fig:a_concentration} at four different $t$ values. Note the increased length of the axon as an effect of the three times larger advection velocity $a$. Note also the sharper gradient close to the growth cone, cf.\ \citet[Sec.~4]{SDJTB1}.}}
\label{fig:solution_c_slices_at_t_s_big_a}
\end{figure}

{
\section{Two-dimensional slices of Figures \ref{fig:solution_c} and \ref{fig:a_concentration}} \label{app:figure_slices}

We complement the three-dimensional plots (Figs. \ref{fig:solution_c} and \ref{fig:a_concentration}) with two-dimensional slices at different $x$ and $t$ values. Recall that in Fig. \ref{fig:solution_c} the tubulin concentration along the axon is plotted for nominal values on the biological and physical parameters. For Fig. \ref{fig:a_concentration} a three times larger advection velocity $a$ is used. The slices of \ref{fig:solution_c} and \ref{fig:a_concentration} are presented next to each other in Figs. \ref{fig:solution_c_slices_at_x_s}--\ref{fig:solution_c_slices_at_t_s_big_a} for easy comparison.
}

\bibliographystyle{spbasicemph}      
\bibliography{ref}   


\end{document}